# Early stages of dissolution corrosion in 316L and DIN 1.4970 austenitic stainless steels with and without anticorrosion coatings in static liquid lead-bismuth eutectic (LBE) at 500°C

Evangelia Charalampopoulou*[1,2], Konstantina Lambrinou*[1,3], Tom Van der Donck[4], Boris Paladino[5], Fabio Di Fonzo[5], Clio Azina[6], Per Eklund[6], Stanislav Mráz[7], Jochen M. Schneider[7], Dominique Schryvers[2], Rémi Delville[1]

[1] SCK CEN, Boeretang 200, 2400 Mol, Belgium

[2] Electron Microscopy for Materials Science (EMAT), University of Antwerp, Groenenborgerlaan 171, 2020 Antwerp, Belgium

[3] School of Computing and Engineering, University of Huddersfield, Queensgate, Huddersfield HD1 3DH, UK

[4] Department of Materials Engineering, KU Leuven, Kasteelpark Arenberg 44, 3001 Leuven, Belgium

[5] Centre for Nano Science and Technology @PoLiMi, Istituto Italiano di Tecnologia, Via Pascoli 70/3, 20133 Milano, Italy

[6] Thin Film Physics Division, Department of Physics (IFM), Linköping University, 58183 Linköping, Sweden

[7] Materials Chemistry, RWTH Aachen University, Kopernikusstr. 10, 52074 Aachen, Germany

**Abstract:**

This work addresses the early stages (≤1000 h) of the dissolution corrosion behavior of 316L and DIN 1.4970 austenitic stainless steels in contact with oxygen-poor ($C_0 < 10^{-8}$ mass%), static liquid lead-bismuth eutectic (LBE) at 500°C for 600-1000 h. The objective of this study was to determine the relative early-stage resistance of the uncoated steels to dissolution corrosion and to assess the protectiveness of select candidate coatings ($Cr_2AlC$, $Al_2O_3$, $V_2Al_xC_y$). The simultaneous exposure of steels with intended differences in microstructure and thermomechanical state showed the effects of steel grain size, density of annealing/deformation twins, and secondary precipitates on the steel dissolution corrosion behavior. The findings of this study provide recommendations on steel manufacturing with the aim of using the steels to construct Gen-IV lead-cooled fast reactors.

# 1  Introduction

A prerequisite for the deployment of Gen-IV lead-cooled fast reactors (Gen-IV LFRs) is the good compatibility of structural and fuel cladding steels with the heavy liquid metal (HLM) coolant(s), such as lead (Pb) and lead-bismuth eutectic (LBE), so as to prevent in-service steel degradation due to liquid metal corrosion (LMC). Depending on the exposure conditions (i.e., temperature, HLM oxygen concentration, HLM flow velocity, exposure duration, temperature gradients, etc.) and the specific steel/HLM pair, the observed LMC effects are characteristic of steel oxidation, dissolution, pitting, erosion, or a combination thereof [1-8]; all standard LMC effects may be exacerbated by liquid metal embrittlement, fretting wear, radiation embrittlement, etc. Moreover, the compositional and microstructural steel homogeneity as well as the steel thermomechanical state (e.g., degree of plastic deformation, steel deformation temperature, etc.) affect greatly the prevalent LMC phenomena and their severity for a specific set of exposure conditions [9,10]. Quite importantly, prior studies have demonstrated that the HLM ingress into the steel is facilitated by specific steel microstructural features, such as grain and deformation twin boundaries; interfaces between steel matrix and secondary precipitates (e.g., carbides, sulfides, oxides, δ-ferrite in austenitic stainless steels); periodic inhomogeneities in chemical composition (also known as 'chemical banding'); etc. [3,8,11-13]. The mitigation of undesirable LMC effects in HLM-cooled nuclear reactors, such as the accelerator-driven system (ADS) system MYRRHA (multi-purpose hybrid research reactor for high-tech applications [14]), relies on the judicious combination of moderate operation temperatures, active oxygen control, in-depth understanding of the LMC mechanisms and, whenever needed, possible implementation of steel surface protection approaches (e.g., PLD-deposited alumina [15-19], GESA surface modification [20]). Active oxygen control aims at steel passivation (i.e., covering the steel surface with thin, protective oxide scales) by maintaining the dissolved oxygen concentration in the HLM coolant, $C_O$, within appropriate, system-specific limits (these limits depend on the $\Delta T$ of the reactor system, the steels used for its construction, the specific HLM coolant, etc.) [21-26].

In case the formation of a protective oxide scale is suppressed in specific reactor locations or its integrity is locally jeopardized (due to mechanical or chemical reasons), dissolution corrosion occurs, whereupon the dissolution of steel alloying elements (e.g., Ni, Mn, Cr, Fe) in the HLM is accompanied by the progressively deeper, with time, HLM penetration into the steel. The depth of HLM dissolution attack can either be uniform over large steel surface areas, or locally-enhanced, leading to dissolution 'pitting' [8]. The latter is particularly undesirable, as it might jeopardize the integrity of thin-walled components, such as fuel cladding tubes and heat exchanger tubes, within very short periods of time. During isothermal steel exposures to oxygen-depleted static HLMs (i.e., conditions promoting dissolution corrosion), diffusion determines the transport of steel alloying elements directly from the solid steel into the HLM, through a diffusion boundary layer that is established at the steel/HLM interface [5]. Ideally, when the HLM becomes saturated in a specific element (i.e., when the temperature-dependent solubility limit of that element is reached), the dissolution of this particular element is expected to eventually stop. Obviously, the elemental solubilities in the HLM [27,28] and elemental diffusion coefficients in the steel depend on temperature, resulting in progressively higher dissolution rates as the temperature increases. Moreover, in a non-isothermal system, steel alloying elements dissolving in the 'hot leg' of the system might precipitate in the 'cold leg', as elemental solubility is temperature-dependent, complicating the interpretation of corrosion phenomena [28,29].

Even though active oxygen control is a theoretically sound LMC mitigation approach, it cannot provide complete shielding from undesirable LMC effects in a real nuclear system. Various reasons (e.g., high-temperature transients, oxygen-depleted localized zones of static LBE, mechanical or chemical degradation of the passivating oxides, non-oxidizing steel inclusions close to the steel surface, etc.) may trigger the occurrence of dissolution corrosion. It is imperative, therefore, to understand the dissolution corrosion mechanism of the candidate (structural and fuel cladding) steel grades considered for the construction of a particular LBE-cooled nuclear system, such as MYRRHA. Realizing that dissolution corrosion may start almost immediately at locations where protective oxide scales cannot form due to fabrication-induced steel defects (e.g., non-oxidizing inclusions close to the steel surface) or at locations where the oxide scales are damaged makes the in-depth study of the steel dissolution corrosion behavior imperative for the safe design of the nuclear system. More specifically, it is important to understand the (steel-specific) dissolution corrosion kinetics as function of temperature, so as to determine the corrosion allowances for a particular steel and a specific reactor component. Taking this a step further, such fundamental studies, where the candidate steels are tested under very conservative conditions (static, oxygen-poor HLMs) are needed to even determine the size of the reactor system so as to ensure its safe operation during its intended lifetime.

This work performed a series of experiments targeting the manifestation of dissolution corrosion in specimens made of the MYRRHA candidate austenitic stainless steels, i.e., the 316L structural steel and the DIN 1.4970 fuel cladding steel. All steel exposures were performed at very conservative and directly comparable test conditions promoting dissolution corrosion, i.e., at 500°C, in oxygen-poor ($C_0 << 10^{-8}$ mass%), static liquid LBE; all tests were designed so as to ensure that the LBE bath would not be saturated in dissolved steel alloying elements during testing, as this would decelerate the progress of LBE dissolution attack. The exposure time was limited (600-1000 h) in order to investigate the early stages of the steel dissolution corrosion behavior. The exposed steels were the two candidate austenitic stainless steels selected for the construction of the MYRRHA system, i.e., the 316L structural steel and the DIN 1.4970 fuel cladding steel. Specimens of both (316L and DIN 1.4970) steel grades, with known microstructural differences and variable degree of cold work, have been exposed simultaneously to liquid LBE in order to assess the effect of various steel microstructural aspects on their resistance to LBE dissolution attack. Moreover, uncoated and coated DIN 1.4970 fuel cladding tube specimens were simultaneously exposed to liquid LBE, so as to compare the effectiveness of promising anticorrosion candidate coatings as perspective dissolution corrosion mitigation strategies for LBE-cooled nuclear reactors.

## 2 Experimental

### 2.1 Test setups & exposure conditions

Three exposures (tests 1, 2 and 3 in Fig. 1) were designed and performed to study the effect of the steel microstructure and thermomechanical state on the early stages (~600 h) of dissolution corrosion of *uncoated* specimens made of the MYRRHA candidate austenitic stainless steels 316L (structural steel) and DIN 1.4970 (fuel cladding steel) in contact with oxygen-poor ($C_O \ll 10^{-8}$ mass%), static liquid LBE at the elevated temperature of 500°C. A schematic representation of the setups used to test uncoated 316L and DIN 1.4970 steels is shown in Fig. 1. Another exposure (test 4 in Fig. 3a) was performed to test the protectiveness of candidate ceramic coatings on the DIN 1.4970 MYRRHA fuel cladding steel under conditions promoting dissolution corrosion; in that test, both *uncoated and coated* DIN 1.4970 fuel cladding tube segments were exposed for 1000 h to oxygen-poor ($C_O < 10^{-10}$ mass%), static liquid LBE at 500°C. A schematic representation of the setup used to expose uncoated and coated DIN 1.4970 cladding steels is shown in Fig. 3a.

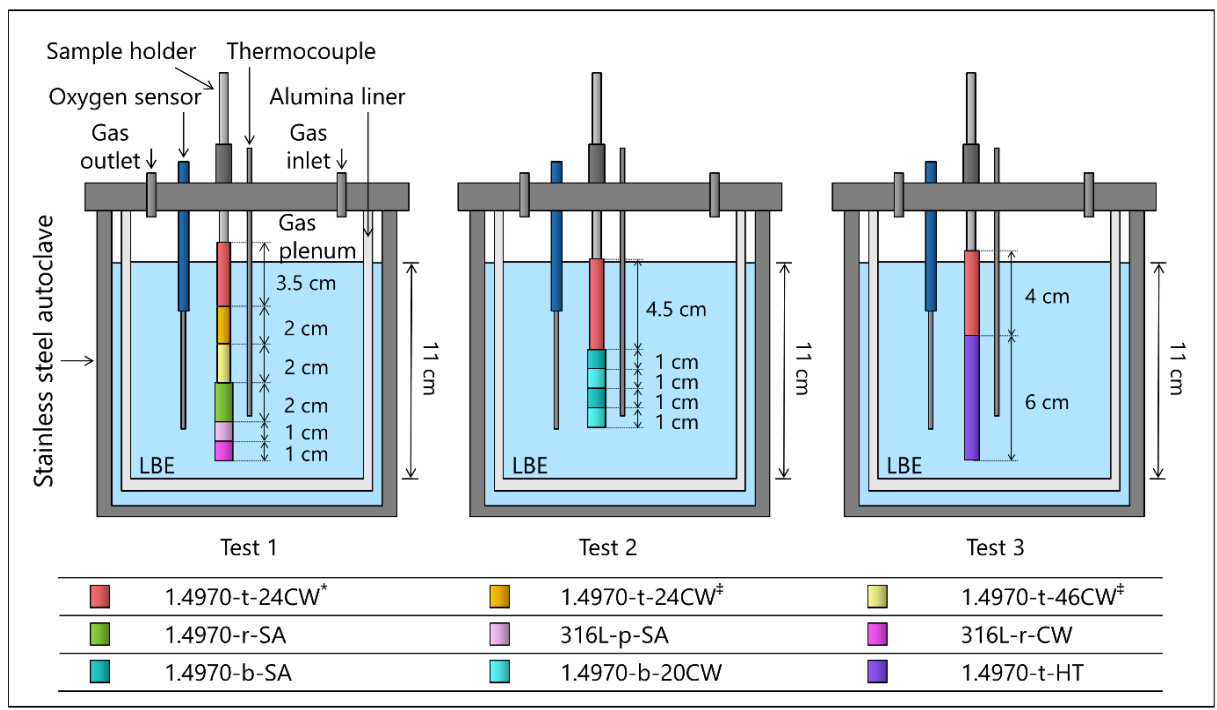

**Fig. 1.** Experimental setups used to expose uncoated 316L and DIN 1.4970 steels to oxygen-poor, static liquid LBE at 500°C for ~600 h. Openings in the lid of the steel autoclave allow the insertion of the sample holder, oxygen sensor, thermocouple, and gas inlet/outlet supply. An alumina liner prevents the contact of the LBE bath with the steel autoclave walls. The color-coded legend allows the distinction of the different steel specimens in each test setup. Symbol * refers to a 'brushed & polished' cladding tube surface state; symbol ‡ refers to a 'brushed' cladding tube surface state.

The test setups for all exposures were identical stainless steel autoclaves with inner alumina (Al$_2$O$_3$) liners to prevent the direct contact of the liquid LBE bath (i.e., the LBE volume in contact with the steel specimens) with the autoclave walls. A detailed description of the prototype test setup (schematically depicted in Figs. 1 and 3a) may be found elsewhere [8]. The temperature and LBE oxygen concentration during the exposures were monitored by thermocouples type K and electrochemical oxygen sensors (i.e., Bi/Bi$_2$O$_3$ reference electrode in test 1 [30]; air/lanthanum strontium manganese oxide (LSM) reference electrode in tests 2-4 [31], respectively. Each test used ~5 kg of 'fresh' liquid LBE, the exact composition and impurities of which are reported elsewhere [8]. In all tests, the exposure temperature was kept constant at ±2°C from the targeted temperature of 500°C, while the concentration of dissolved oxygen in liquid LBE was maintained low ($C_O$ << 10$^{-8}$ mass%) by the continuous purging with a reducing conditioning gas (HYTEC 5: Ar-5% H$_2$, Rapid Industrial Gases Ltd., UK). The LBE oxygen concentration and temperature profiles during the LBE exposures of uncoated 316L and DIN 14970 steels (tests 1-3 in Fig. 1) is shown in Fig. 2, while the same profiles during the LBE exposure of uncoated and coated DIN 1.4970 steels (test 4 in Fig. 3a) is shown in Fig. 3b.

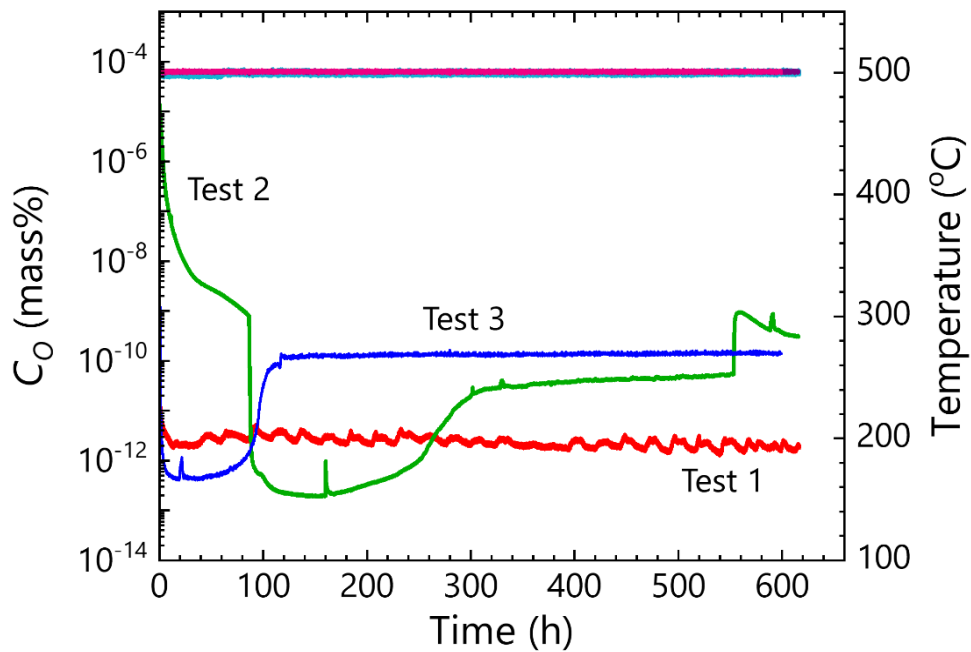

**Fig. 2.** Temperature and LBE oxygen concentration profiles during the exposures of uncoated 316L and DIN 1.4970 steels at 500°C for ~600 h (tests 1, 2, 3 in Fig. 1).

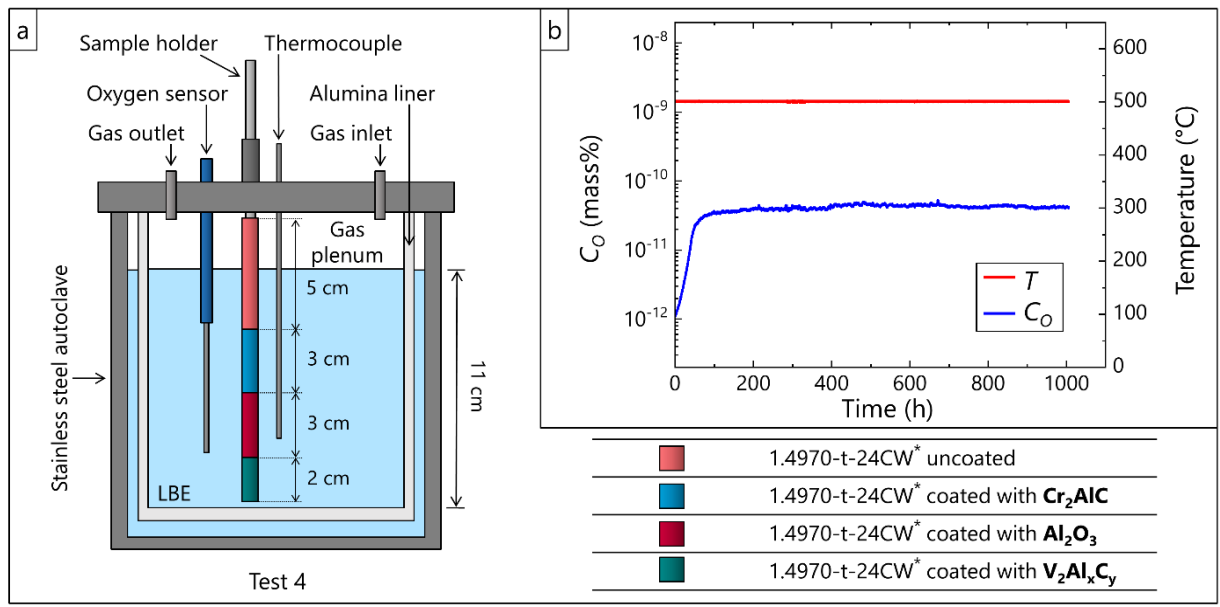

**Fig. 3.** (a) Experimental setup used to expose uncoated and coated DIN 1.4970 cladding tube segments to oxygen-poor, static liquid LBE at 500°C for 1000 h. (b) Temperature and LBE oxygen concentration profile during the exposure of uncoated and coated DIN 1.4970 fuel cladding tube segments. The color-coded legend describes the exposed DIN 1.4970 stainless steel specimens.

*2.2    Test rationale & exposed materials*

The primary aim of this work was to compare the early stages of the dissolution corrosion behaviour of the *uncoated* MYRRHA austenitic stainless steels 316L (structural components), and DIN 1.4970 (fuel cladding tubes). Therefore, uncoated specimens of two different 316L steel heats, i.e., one solution-annealed (316L-p-SA) and one cold-worked (316L-r-CW), were exposed to oxygen-poor, static liquid LBE in test 1 together with uncoated DIN 1.4970 steel specimens (bulk and tube specimens, different degrees of cold work, different surface finishes). It must be pointed out that all DIN 1.4970 steels (rod, bars, tubes), as well as the specimens machined out of them and tested in this work (tests 1-4), were fabricated from the same starting steel heat. Table 1 provides the chemical composition of the three different steel heats (316L-p-SA, 316L-r-CW, DIN 1.4970) used in this work.

**Table 1.** Elemental composition (in mass%) of the DIN 1.4970, 316L SA & 316L CW steel heats.

|  | Cr | Ni | Mo | Mn | Si | Ti | P | C | S | Fe |
|---|---|---|---|---|---|---|---|---|---|---|
| DIN 1.4970 | 15.06 | 15.05 | 1.21 | 1.86 | 0.57 | 0.44 | 0.013 | 0.096 | <0.001 | Bal. |
| 316L-p-SA | 16.73 | 9.97 | 2.05 | 1.81 | 0.67 | - | 0.032 | 0.019 | 0.004 | Bal. |
| 316L-r-CW | 16.68 | 10.10 | 2.07 | 1.75 | 0.27 | - | 0.041 | 0.027 | 0.028 | Bal. |

At this point, it should be mentioned that the specification of a nuclear fuel cladding steel is very strict, imposing a precise and well-controlled fabrication process, due to the key role of such material in the safe and efficient operation of the nuclear reactor. The reference MYRRHA DIN 1.4970 fuel cladding steel, fabricated by Sandvik, Sweden, is a Ti-stabilized 1.4970 '15-15Ti' austenitic stainless steel; in this work, the "standard" version of the DIN 1.4970 fuel cladding steel is 24% cold-worked (CW) with a 'brushed and polished' surface finish (herein denoted as 1.4970-t-24CW). The DIN 1.4970 stainless steel grade was selected as the MYRRHA reference fuel cladding, based on the extensive database of material properties (including performance under neutron irradiation) for Gen-IV sodium-cooled fast reactors (SFRs). The DIN 1.4970 steel grade contains a large number of stabilizing Ti(C,N) precipitates, which have a beneficial effect on its thermal creep properties by limiting dislocation motion [32]. Due to the limited understanding of the dissolution corrosion behaviour of the DIN 1.4970 stainless steel grade, not only in view of the more extensively investigated 316L steel grade, but also in terms of the effect of various microstructural aspects (grain size, degree of cold work, etc.) on its corrosion behaviour, tests 1-3 have exposed different uncoated DIN 1.4970 steel specimens to liquid LBE.

Test 1 (500°C, $C_0 < 10^{-9}$ mass%, 619 h) included two 316L steel bulk specimens: (i) one machined from the solution-annealed heat 316L-p-SA ($\varnothing$ 9 mm, 1 cm) and (ii) one machined from the cold-worked heat 316L-r-CW ($\varnothing$ 9 mm, 1 cm). Test 1 also included four DIN 1.4970 steel specimens: (i-ii) two 1.4970-t-24CW tubes (outer $\varnothing$ 6.55 mm, inner $\varnothing$ 5.65 mm; one 3.5 cm-long in the 'brushed & polished' (*) surface state, and one 2 cm-long in the 'brushed' (‡) surface state) with 24% CW, (iii) one 1.4970-t-46CW tube (outer $\varnothing$ 6.55 mm, inner $\varnothing$ 5.65 mm, 2 cm; brushed surface state) with 46% CW, and (iv) one bulk specimen ($\varnothing$ 9 mm, 2 cm) machined from the solution-annealed rod 1.4970-r-SA. Pictures of the two different 1.4970-t-24CW surfaces finishes (brushed & polished, brushed) are shown in Fig. S1 (supplementary information). All bulk steel specimens have been polished with down to 3 μm diamond paste, cleaned with acetone and ethanol in an ultrasonic bath, and blow-dried before their insertion to the LBE bath [8].

Test 2 (500°C, $C_0 < 10^{-8}$ mass%, 613 h) included five DIN 1.4970 steel specimens: (i) one 1.4970-t-24CW tube (outer $\varnothing$ 6.55 mm, inner $\varnothing$ 5.65 mm, 4.5 cm; brushed & polished surface state), (ii-iii) two bulk specimens ($\varnothing$ 13 mm, 1 cm) machined from the solution-annealed bar 1.4970-b-SA, and (iv-v) two bulk specimens ($\varnothing$ 13 mm, 1 cm) machined from the 20% CW bar 1.4970-b-20CW. All four bulk specimens were inserted in the LBE bath with a polished surface similar to that of the bulk specimens exposed in test 1.

Test 3 (500°C, $C_0 < 10^{-10}$ mass%, 600 h) included two tube DIN 1.4970 steel specimens: (i) one (as-received) 1.4970-t-24CW tube (outer $\varnothing$ 6.55 mm, inner $\varnothing$ 5.65 mm, 4 cm; brushed & polished surface state), and (ii) one heat-treated 1.4970-t-HT tube (outer $\varnothing$ 6.55 mm, inner $\varnothing$ 5.65 mm, 6 cm; 'as-annealed' surface state). In order to protect the DIN 1.4970 tube segment during the annealing heat treatment (1000°C, 2 h), the tube specimen was vacuum-encapsulated in a quartz (SiO$_2$) tube together with a zirconium (Zr) tube and a tantalum (Ta) foil, which were used as oxygen getters (see Fig. S2, supplementary information). The annealed DIN 1.4970 tube did not show any visible signs of oxidation prior to its immersion in the LBE bath.

Table 2. Description of uncoated 316L and DIN 1.4970 steels exposed to liquid LBE in tests 1-3; CW: cold-worked, SA: solution-annealed, HT: heat treated; t: tube, b: bar, r: rod, p: plate.

| Steel grade | Thermomechanical state | Starting steel geometry | Code name | Specimen surface state |
|---|---|---|---|---|
| DIN 1.4970 | 24% CW | Tube | 1.4970-t-24CW*; 1.4970-t-24CW‡ | brushed & polished (*); brushed (‡) |
| | 46% CW | Tube | 1.4970-t-46CW‡ | brushed (‡) |
| | HT | Tube | 1.4970-t-HT | as-annealed |
| | 20% CW | Bar | 1.4970-b-20CW | polished |
| | SA | Bar | 1.4970-b-SA | polished |
| | SA | Rod | 1.4970-r-SA | polished |
| 316L | CW | Rod | 316L-r-CW | polished |
| | SA | Plate | 316L-p-SA | polished |

Tests 1-3 included 1.4970-t-24CW tube segments of the MYRRHA reference fuel cladding material, so as to benchmark the resistance of all DIN 1.4970 steels specimens (with differences in the degree of cold work, microstructure, surface state, etc.) to dissolution corrosion under directly comparable exposure conditions. Table 2 describes the uncoated 316L and DIN 1.4970 steels exposed to liquid LBE in tests 1-3, indicating the differences in thermomechanical state, starting steel geometry, and steel surface state prior to LBE exposure. As mentioned earlier, tubes 1.4970-t-24CW and 1.4970-t-46CW, bars 1.4970-b-SA and 1.4970-b-20CW, and rod 1.4970-r-SA exposed in tests 1-3 came from the same steel heat; more information on the processing route of the DIN 1.4970 stainless steel used in this work is provided in section 2.2.2. Code names have been assigned to distinguish the steel specimens; these code names will be henceforth used to discuss the findings of this work.

A secondary objective of this work was to assess the protectiveness of different anticorrosion coatings on the MYRRHA reference DIN 1.4970 fuel cladding steel (1.4970-t-24CW; brushed and polished surface). Two of the coatings were rather innovative, targeting the deposition of MAX phase compounds, i.e., Cr$_2$AlC and V$_2$AlC, on the substrate DIN 1.4970 fuel cladding steel. The MAX phases are ternary carbides and nitrides described by the $M_{n+1}AX_n$ general stoichiometry, where M is an early transition metal, A is an A-group element (commonly group 13 or 14), X is C or N, and n = 1, 2, 3 [33]. The MAX phases have shown excellent compatibility with HLMs [34-36] as well as superior radiation tolerance above 600°C [37-40], making them excellent candidate anticorrosion coatings for Gen-IV LFR fuel cladding steels. The third coating targeting the deposition of alumina ($Al_2O_3$), an oxide known for its stability at very low oxygen potentials in both liquid Pb and LBE [23], making an excellent coating material capable of preventing the HLM dissolution attack of the substrate steel. The alumina protectiveness is the cornerstone of many state-of-the-art LMC mitigation approaches, such as the deposition of $Al_2O_3$ coatings [15-19], the GESA steel surface modification (usually by means of a FeCrAl(Y) top layer), and the development of alumina-forming austenites (AFAs) [41] and/or aluminium-containing high-entropy alloys (HEAs) [42]. The concurrent exposure of (standard) alumina-coated and (innovative) MAX phase-coated DIN 1.4970 fuel cladding tubes to oxygen-poor, static liquid LBE aimed at providing a first assessment of the relative resistance of these different coating materials to dissolution corrosion. The targeted coating deposition temperature for all coatings made in this work was below 600°C, so as to limit the possible degradation of the bulk properties of the DIN 1.4970 substrate fuel clad by high coating deposition temperatures. The coatings were deposited on different segments of the same 1.4970-t-24CW fuel cladding tube, so as to allow for a direct comparison of the performance of the three different coatings on the exact same substrate clad. Detailed description of the actual deposition processes for the three anticorrosion coatings on DIN 1.4970 fuel cladding tubes that were tested in this work is provided in the supplementary information.

Test 4 (500°C, $C_0 < 5\times10^{-11}$ mass%, 1000 h) included four tube DIN 1.4970 steel specimens (1.4970-t-24CW; outer ⌀ 6.55 mm, inner ⌀ 5.65 mm; brushed & polished surface): (i) one uncoated (5 cm-long), (ii) one coated by magnetron sputtering with the $Cr_2AlC$ MAX phase (coating thickness ∼3 μm; 3 cm-long), (iii) one coated by pulsed laser deposition (PLD) with $Al_2O_3$ (coating thickness ∼7 μm; 3 cm-long), and (iv) one coated by cathodic arc deposition with $V_2Al_xC_y$ (coating thickness < 200 nm; 2 cm-long). Table 3 provides information on the anticorrosion coatings deposited on 1.4970-t-24CW fuel cladding tube segments. The $V_2Al_xC_y$ coating was initially intended as a $V_2AlC$ coating, however, this MAX phase compound did not form at the low deposition temperature (500-550°C).

**Table 3.** Description of anticorrosion coatings on DIN 1.4970 fuel cladding tube segments (1.4970-t-24CW, brushed & polished surface) exposed to liquid LBE in test 4.

| Substrate steel | Coating | | Coating deposition | |
| --- | --- | --- | --- | --- |
| | Composition | Thickness | Deposition method | Deposition temperature (°C) |
| DIN 1.4970 (1.4970-t-24CW) | $Cr_2AlC$ | ∼3 μm | Magnetron sputtering | 580 |
| | $Al_2O_3$ | ∼7 μm | Pulsed laser deposition | RT† |
| | $V_2Al_xC_y$ | <200 nm | Cathodic arc deposition | 500-550 |

† RT = room temperature

### 2.3 316L & DIN 1.4970 stainless steels

#### 2.3.1 316L stainless steel heats

This work exposed simultaneously to liquid LBE specimens machined from the following two heats of the 316L MYRRHA candidate structural steel: (a) a solution-annealed (1050-1100°C), cold rolled plate of industrial size (15 mm in thickness; EUROTRANS-DEMETRA heat; Industeel, ArcelorMittal, S.A. [43]), which is herein named 316L-p-SA (Table 2); and (b) a cold-drawn steel heat (cylindrical rod of ∅ 10 mm; SIDERO STAAL nv, Belgium), which is herein named 316L-r-CW (Table 2). The simultaneous exposure of different 316L stainless steel heats to liquid LBE (test 1, Fig. 1) allows the assessment of the steel microstructure and thermomechanical state on the steel corrosion behaviour at the selected test conditions, making the obtained test results grade-specific rather than heat-specific, as previously demonstrated [8]. Lambrinou et al. [8] had previously exposed the same two 316L steel heats to conditions favouring dissolution corrosion (i.e., 500°C, $C_0$ < $10^{-8}$ mass%, 253-3282 h); in that earlier work, the 316L-p-SA solution-annealed steel was named 316LSA, while the 316L-r-CW cold-drawn steel was named 316LH2. The present study uses the reliable corrosion data earlier collected on 316L steels [8] to investigate the relative resistance of both MYRRHA candidate steels (i.e., 316L and DIN 1.4970) to dissolution corrosion under exposure conditions (500°C, $C_0$ < $10^{-9}$ mass%, 619 h) similar to those previously used.

#### 2.3.2 DIN 1.4970 stainless steel heat

Both tube and bulk specimens of the DIN 1.4970 MYRRHA candidate fuel cladding steel exposed in this work to liquid LBE were made of the same steel heat produced by Sandvik, Sweden. The fabrication of this DIN 1.4970 heat involved melting of 10 tons of steel in a high frequency induction furnace, followed by melt refinement in a CLU (Creusot Loire Uddeholm) converter, so as to reduce the carbon (C), sulphur (S) and oxygen (O) contents as well as the slag layer [44]. Square blooms of high purity were subjected to several steps of hot rolling and hot forging, first producing the 1.4970-r-SA rod (∅ 121 mm). This rod was characterised by a highly inhomogeneous microstructure with grain size that increased drastically from the centre of the rod to its periphery, and a bimodal grain size distribution throughout the rod volume [45]. The 1.4970-r-SA specimen (∅ 9 mm, 2 cm) exposed to liquid LBE in test 1 was machined from the outside of the rod, where the grains reached several mm in diameter. Exposing to LBE a specimen from the highly coarse-grained part of the 1.4970-r-SA rod aimed at studying the dissolution corrosion behaviour of a DIN 1.4970 steel with limited grain boundary surface per unit volume.

Rod 1.4970-r-SA was hot rolled and cold drawn into bars (∅ 14.5 mm), which were further either cold drawn for a second time followed by annealing (1125°C, 5 min), thus producing the annealed 1.4970-b-SA bar (∅ 13 mm), or were subjected to annealing followed by cold drawing, thus producing the cold-drawn 1.4970-b-20CW bar (∅ 13 mm). Both bars retained a bimodal grain size distribution, which was less pronounced than that observed in the 1.4970-r-SA rod. Microstructural analysis of the central area of both bars revealed that the annealed 1.4970-b-SA bar was finer-grained than the cold-worked 1.4970-b-20CW bar, presumably due to the two consecutive cold drawing steps. The simultaneous exposure of specimens (∅ 13 mm, 1 cm) made from the complete diameter of bars 1.4970-b-SA and 1.4970-b-20CW to liquid LBE in test 2 aimed at comparing milder microstructural differences (in grain size, type and density of defects induced by cold deformation, etc.) on the DIN 1.4970 steel dissolution corrosion behaviour.

The production of DIN 1.4970 fuel cladding tubes started from square blooms that underwent several steps of hot rolling and forging to the appropriate dimensions, thus allowing the hot extrusion of deep-bored cylindrical bar sections. After extrusion, the tube wall thickness was reduced by a cold pilgering process to produce 'hollows', the starting material for the fabrication of tubes (Sandvik-Précitube, France) with very precise final wall thickness (450 μm). Further reduction of the hollows' diameter and wall thickness was ensured by four plug-drawing operations at RT. A short high-temperature annealing treatment was performed between each cold-drawing pass, the last one of which determined the degree of cold work (24% CW) in the 1.4970-t-24CW fuel cladding tubes [44]. Some of the 'hollows' were used to produce highly cold-worked 1.4970-t-46CW tubes, so as to simulate irradiation hardening [46]. 1.4970-t-24CW and 1.4970-t-46CW tubes were simultaneously exposed to liquid LBE in test 1 to compare the effect of the degree of cold work on the steel dissolution corrosion behaviour.

Since it has often been suggested that the steel surface state affects its LMC resistance, two different surface states of the DIN 1.4970 fuel cladding tubes have been exposed to liquid LBE in this work: the standard 'brushed and polished' state, and the less refined 'brushed' surface state. Targeting either of the two surface states during the fabrication route of the reference MYRRHA fuel cladding tubes is feasible on an industrial scale, if one of them showed a better performance during the screening LBE exposures. The average, $R_a$, and maximum, $R_{max}$, surface roughness for the 'brushed and polished' finish were 0.268 μm and 3.65 μm, respectively, while the same values for the 'brushed' finish were 0.322 μm and 3.91 μm, respectively. 1.4970-t-24CW tubes with both surface states were exposed to liquid LBE in test 1 to check whether such minor alterations in the steel surface finish would affect its dissolution corrosion behaviour. In the same test, one 1.4970-t-46CW tube with the 'brushed' surface state was included due to the limited size of the LBE bath.

In order to gain a deeper insight into the dissolution corrosion of the DIN 1.4970 stainless steel, a 1.4970-t-24CW tube was subjected to high-temperature vacuum annealing (1000°C, 2 h). This tube was exposed to liquid LBE in test 3 together with a reference (as-received) 1.4970-t-24CW tube, so as to investigate the effects of high-temperature annealing (e.g., removal of deformation-induced defects, recrystallisation, change in texture) to the resistance of the DIN 1.4970 steel to dissolution corrosion. Knowing the effect of steel microstructural changes, due to possible high-temperature transients, on the steel dissolution corrosion behaviour is of great value for the deployment of Gen-IV LFR nuclear systems.

*2.4   Microstructural characterization of pristine & exposed materials*

Prior to exposing the 316L and DIN 1.4970 stainless steels to liquid LBE, the microstructure of both steels was carefully characterized by light optical microscopy (LOM; Zeiss Axio Scope.A1, ZEISS International), electron backscatter diffraction (EBSD; FEI Scios DualBeam FIB/SEM with Hikari EBSD detector & FEI Nova NanoSEM 450 with Hikari XP EBSD detector), and transmission electron microscopy (TEM; JEM-3010, JEOL). Metallographic cross-sections were prepared from the pristine steels with 1 μm diamond paste in the last step. Some of the polished cross-sections were given an additional polishing step with a colloidal silica (OPS) suspension so as to create a subtle surface relief that allowed better inspection by LOM, while others were chemically etched (Carpenter's etchant: 8.5 g $FeCl_3$, 2.4 g $CuCl_2$, 122 ml HCl, 6 ml $HNO_3$, 122 ml ethanol [8]) to better reveal the grain boundaries (GBs).

After their exposure to LBE, metallographic cross-sections were prepared from all exposed materials. The longer tube specimens were cut in segments, in order to collect information on the (possible) fluctuation of LMC effects along the tube length, by analysing different cross-sections from the bottom, middle, and top of the exposed tubes. All material cross-sections were cold-mounted in non-conductive resin (Struers Specifix-20), polished to a mirror surface finish, and coated with a C thin film for further inspection by means of scanning electron microscopy (SEM; JSM-6610LV, JEOL) and energy-dispersive X-ray spectroscopy (EDS; XFlash detector 4010, Bruker AXS GmbH). Basic characterisation of the exposed steels was conducted by means of LOM, SEM, EDS, and EBSD. The in-depth analysis of select areas of interest demanded the lift-out of thin (electron-transparent) foils by focused ion beam (FIB; FEI Helios Nanolab 650, FEI), so as to allow their subsequent investigation by means of TEM and scanning transmission electron microscopy (STEM; FEI Tecnai Osiris S/TEM, FEI); the latter was equipped with high-angle annular dark field (HAADF) and EDS detectors suitable for Z-contrast and elemental analysis, respectively.

## 3   Results & Discussion

*3.1   Microstructural characterization of pristine (non-exposed) steels*

Apart from the known differences in their chemical composition (Table 1), the pristine 316L and DIN 1.4970 steels showed microstructural differences regarding grain size, type and density of defects (e.g., deformation and annealing twins, secondary precipitates), and crystallographic texture (Figs. 4-7). Identifying the main steel microstructural differences and investigating their defects allowed the interpretation of the observed discrepancies and/or similarities in the early stages of the dissolution corrosion behavior of steels exposed to comparable exposure conditions.

The EBSD orientation maps and corresponding inverse pole figures of Fig. 4 allow for a quick comparison of the grain size and texture of all steels used in this work. The coarser-grained steels appear to be 1.4970-r-SA, 316L-p-SA and 316L-r-CW, followed by 1.4970-b-20CW and 1.4970-b-SA; clearly, the finest-grained steel is 1.4970-t-46CW. Detailed statistical analysis of the EBSD data (acquired from more than one areas in the pristine steels) showed that the weighted average grain diameter of 1.4970-r-SA (i.e., average that considers the area fraction occupied by each grain size) was 205 µm (scan shown in Fig. 4), and that 98% of the scanned area was occupied by grains >10 µm. Actually, the scanned area (286×838 µm$^2$) was too small to capture the largest grains in this part of the 1.4970-r-SA rod, the diameter of which was in the mm range (see section 2.3.2). This was also confirmed by the statistical analysis of the 1.4970-r-SA grain size data from all scans, which indicated a monotonic increase in the area fraction with grain size (i.e., in the scan of Fig. 4, 7.5% of the area was occupied by 107.5 µm-large grains, 10% of the area by 171 µm-large grains, and 61% of the area by 271.5 µm-large grains). The very coarse grains are often characterized by the presence of annealing twins; the fraction of annealing twins in the 1.4970-r-SA scan of Fig. 4 was 44%. The EBSD grain orientation maps shown in Fig. 4 are parts of similarly sized scans, which were sufficiently large to determine the grain size distribution in all steels, except from the overall coarse-grained 1.4970-r-SA and specific coarse-grained areas in 316L-r-CW. The weighted average grain diameter in 316L-p-SA varied between 44 and 51 µm; for an average grain diameter of 47 µm (scan in Fig. 4), 98% of the scanned area was occupied by grains >10 µm, the most commonly occurring (16%) grain size was 42 µm, and the fraction of annealing twins was 21%. The weighted average grain diameter in 316L-r-CW varied in the 68-188 µm range; the average grain diameter in the coarse-grained area of Fig. 4 was 188 µm (scanned area not large enough to capture the largest grains), 96% of the scanned area was occupied by grains >10 µm, and the fraction of annealing twins was 39%. The weighted average grain diameter in 1.4970-b-SA and 1.4970-b-20CW (scans in Fig. 4) was determined to be 18 µm and 30 µm, respectively; this agreed with the initial characterization of the as-received bars (see section 2.3.2), which indicated that the solution-annealed bar was finer-grained than the cold-worked one, due to the two consecutive cold drawing steps during the processing of 1.4970-b-SA. In Fig. 4, 71% of the scanned 1.4970-b-SA area was occupied by grains >10 µm, the most common (11%) grain size was 16 µm, and the fraction of annealing twins was 13%. Moreover, 96% of the scanned 1.4970-b-20CW area was occupied by grains >10 µm, the most common (15%) grain size was 32 µm, and the fraction of annealing twins was 60%. The scans of the two 1.4970 bars were taken next to the specimen surface, hence, they related directly to the observed steel dissolution corrosion damages. The weighted average grain diameter in 1.4970-t-HT (scan in Fig. 4) was 27 µm, 93% of the scanned area was occupied by grains >10 µm, and the fraction of annealing twins was 47%. Last but not least, the weighted average grain diameter in 1.4970-t-24CW and 1.4970-t-46CW (scans in Fig. 4) was determined to be 12 µm and 5.5 µm, respectively; the variations in average grain size in the fuel cladding tubes were very limited due to the strict processing route followed for their production (e.g., the average grain size in the 'reference' 1.4970-t-24CW fuel clad varied in the 11-12 µm range). In 1.4970-t-24CW (Fig. 4), 52% of the scanned area was occupied by grains >10 µm, the most common (12%) grain size was 9.5 µm, and the fraction of annealing twins was 5%.

On the other hand, only 11% of the 1.4970-t-46CW scanned area was occupied by grains >10 μm, the most common (7.5%) grain size was 6.5 μm, and the fraction of annealing twins was 2% (Fig. 4). Finally, it is worthwhile mentioning that reducing the scanned area in 1.4970-t-24CW from 221×480 μm² (scan step: 1 μm) to 93×539 μm² (scan step: 0.5 μm) increased the fraction of twins from 5% to 9%. Similarly, reducing the scanned area in 1.4970-t-46CW from 173×490 μm² (scan step: 1 μm) to 73.5×216 μm² (scan step: 0.5 μm) increased the fraction of twins from 2% to 4%. The increase in the determined twin fraction values for the two fuel cladding steels can be associated with the pickup of finer twins as result of the increase in the EBSD scan resolution (i.e., by reducing the step size used to acquire the scans).

As indicated by the inverse pole figures in Fig. 4, the 316L and DIN 1.4970 steels used in this work are characterized by different crystallographic textures. All cold-worked steels (1.4970-t-24CW, 1.4970-t-46CW, 1.4970-b-20CW, and 316L-r-CW) showed a mixed <111> and <001> texture, typical for austenitic (*fcc*) stainless steels [9,47], such as the 316L and DIN 1.4970 steels used in this work. A less strong texture was observed in all solution-annealed steels (1.4970-t-HT, 1.4970-b-SA, 1.4970-r-SA, and 316L-p-SA), as expected. Even though its effect on the dissolution corrosion behavior of austenitic stainless steels has never been systematically studied, texture can affect locally the progress of LBE dissolution attack. The effect of crystallographic orientation on the (local) dissolution corrosion behavior 316L steels exposed to oxygen-poor, static liquid LBE has been touched upon in prior studies [9,13]. In fact, it has been shown that steel grains with the <111> orientation close to the direction of steel tensile loading (e.g., load resulting in cold deformation) favor LBE dissolution attack, while grains with the <001> orientation delay locally the attack [9]. The higher susceptibility of grains with the <111> orientation to LBE dissolution attack may be attributed to their high density of deformation twins, as such twins are known to promote the LBE ingress into the steel bulk [8,9,13,47]. As was pointed out in earlier studies [48], twinning at low strain levels (in the present study, the degree of steel cold work varied between 20% and 46%) depends strongly on the crystallographic orientation and it is favored in grains with the <111> orientation close to the direction of tensile loading. Table 4 summarizes the fractions of the scanned areas that were occupied by grains with diameters > 10 μm and < 4 μm, the weighted average grain diameters (in μm), and the texture for all exposed steels.

Light optical microscopy has been used in this work to reveal steel microstructural defects (e.g., grain boundaries, annealing/deformation twins, precipitates) that promote dissolution corrosion. Grain boundaries (GBs) and annealing/deformation twins are visible in chemically etched 1.4970-t-24CW, along with a finer grain size distribution close to the (inner/outer) tube surfaces (Fig. 5a). The finer-grained steel zones (10-20 μm-wide) next to the inner/outer cladding tube surfaces are also characterized by the absence of the coarser annealing twins that are otherwise homogenously distributed throughout the wall thickness. The finer steel grains and absence of annealing twins close to the inner/outer cladding tube surfaces, even though not reported previously [44], may be associated with the fuel cladding tube fabrication process. Since grain boundaries are favorable paths for the LBE penetration into the steel bulk [8], knowing the grain size distribution and its inhomogeneities is key for interpreting the dissolution corrosion behavior of a particular steel.

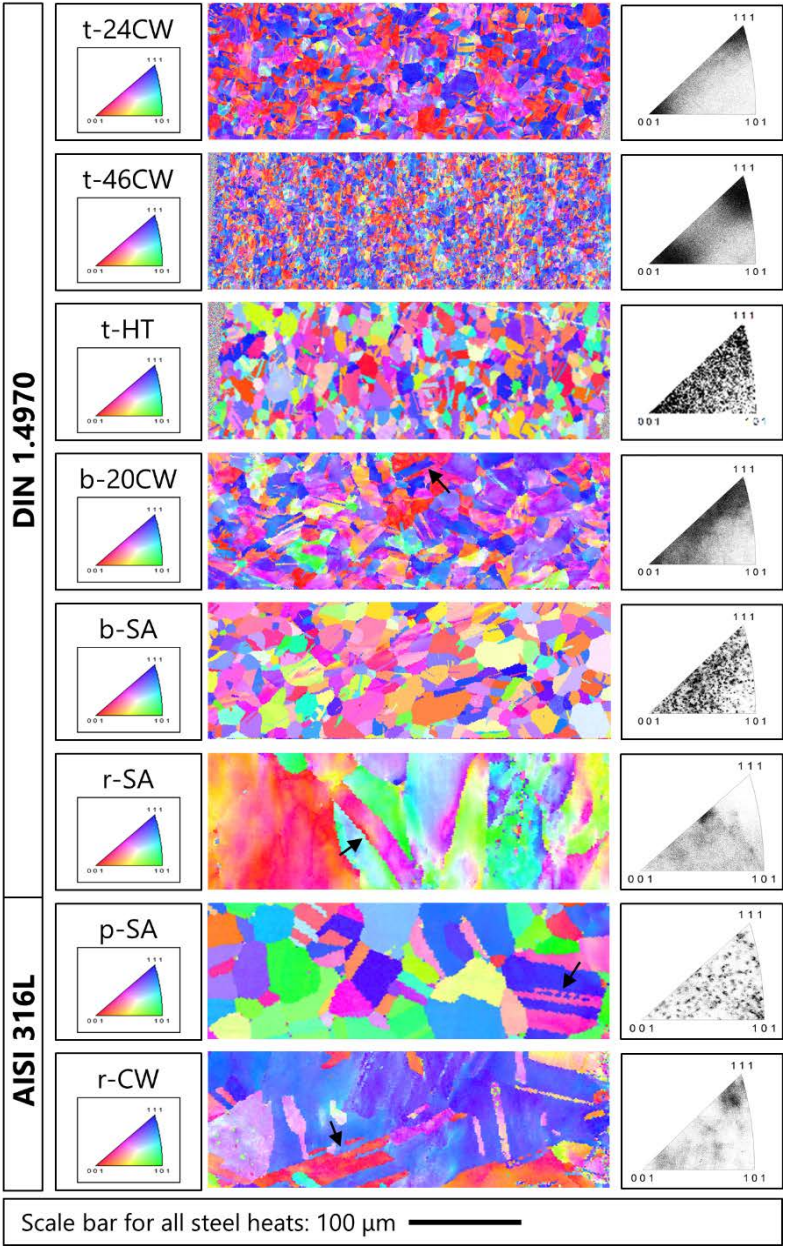

**Fig. 4.** EBSD orientation maps of the 316L and DIN 1.4970 stainless steels exposed to liquid LBE in this work, and corresponding inverse pole figures. The inverse pole figure colour maps correspond to the [001] zone axis. In all orientation maps, going from the left to the right indicates transition from the specimen bulk to its surface. Annealing twins are pin-pointed by arrows in the coarser-grained steels.

**Table 4.** Summary of the fraction (%) of the scanned areas occupied by grain diameters > 10 μm and < 4 μm, the weighted average grain diameter (μm), and the texture of all exposed steels.

| Steel grade | Code name | Scanned area with grains >10 μm (%) | Scanned area with grains <4 μm (%) | Weighted average grain diameter (μm) | Texture |
|---|---|---|---|---|---|
| DIN 1.4970 | 1.4970-t-24CW | 52 | 10 | 12 | <111> & <001> |
| | 1.4970-t-46CW | 11 | 38 | 5.5 | <111> & <001> |
| | 1.4970-t-HT | 93 | 1 | 27 | weak |
| | 1.4970-b-20CW | 96 | 1 | 30 | <111> & <001> |
| | 1.4970-b-SA | 71 | 7 | 18 | weak |
| | 1.4970-r-SA | 98 | 1 | 205 | <111> & <001> |
| 316L | 316L-r-CW | 96 | 1 | 188 | <111> & <001> |
| | 316L-p-SA | 98 | 0.4 | 47 | weak |

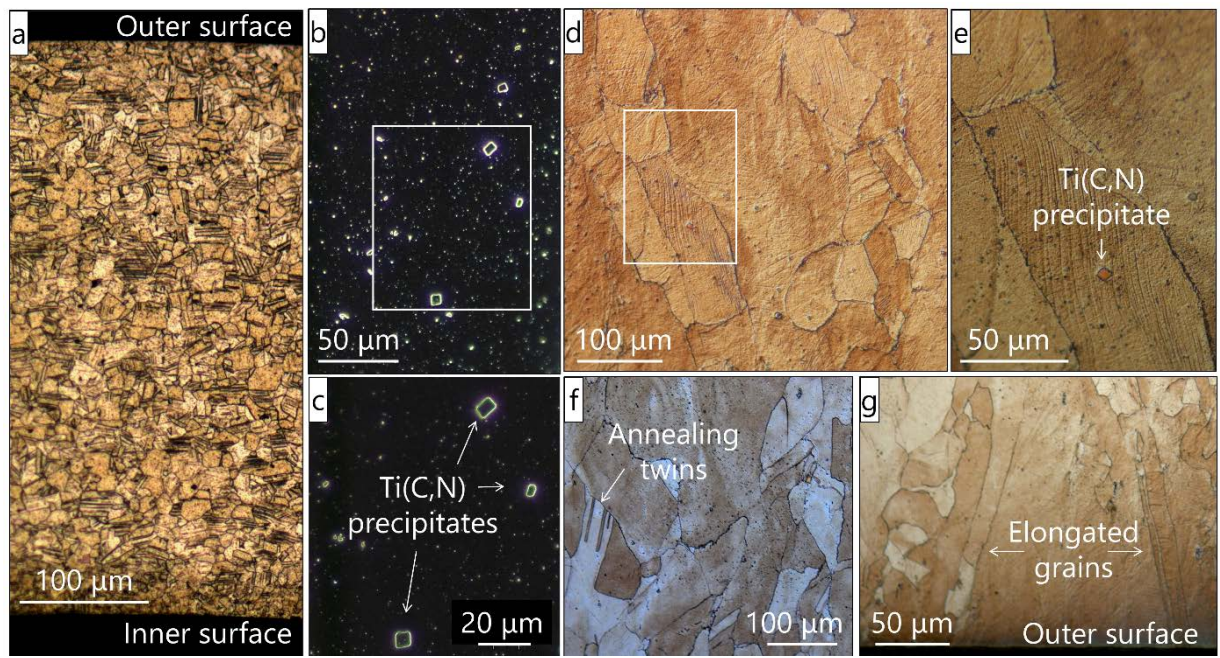

**Fig. 5.** LOM images of DIN 1.4970 steels: (a-c) 1.4970-t-24CW, and (d-g) 1.4970-r-SA. (a) Overview of chemically etched 1.4970-t-24CW. (b,c) LOM dark field images (overview & detail) of the distribution of Ti(C,N) precipitates in OPS-polished 1.4970-t-24CW. (d-f) Bimodal grain distribution in chemically-etched 1.4970-r-SA; microstructural features, such as Ti(C,N) precipitates, GBs, and annealing twins, are visible. (g) Large (>150 μm) elongated grains at high angles (70-80°) with the outer surface of the 1.4970-r-SA specimen.

Dark field LOM images of OPS-polished 1.4970-t-24CW revealed a bimodal distribution of Ti(C,N) precipitates across the wall thickness (Figs. 5b-5c). With the exception of a few, very large (4.5-9 µm) Ti(C,N) precipitates, the rest of the precipitates appeared to be in the submicrometer range. Prior characterization of the MYRRHA reference 1.4970-t-24CW fuel cladding showed that the submicrometer Ti(C,N) precipitates could be divided into 'primary' (50-300 nm) and 'secondary' (1-20 nm) ones [44]. The ultra-fine 'secondary' precipitates are held accountable for the superior steel resistance to (thermal & radiation) creep and radiation swelling [44]. Knowing the distribution of foreign steel precipitates can provide additional insights into the dissolution corrosion behavior of a specific steel. For example, Lambrinou et al. [8] demonstrated that such precipitates (oxides, sulphides, δ-ferrite) in 316L stainless steels promote the local advancement of the dissolution front by allowing the faster LBE penetration into the steel/precipitate interface.

The LOM images in Figs. 5d-5g reveal 1.4970-r-SA microstructural features that could affect the steel dissolution corrosion behavior: bimodal grain and Ti(C,N) precipitates size distributions, large annealing twins, and elongated (>150 µm) grains at high angles with the outer specimen surface. The latter is considered potentially problematic, as GBs are preferred paths of LBE ingress into the steel and having such paths that run deep (>150 µm) into the steel might result in the formation of dissolution 'pits'. Lambrinou et al. [8] has previously reported dissolution 'pitting' in 316L steels due to the unfavorable orientation and inward convergence of large annealing twin boundaries (another defect favoring LBE penetration) with access to the steel specimen surface.

As annealing twins are steel microstructural features that have previously shown that they are effective in aiding locally the progress of LBE dissolution attack and are even capable of forming dissolution 'pits' [8], it has been deemed necessary in this work to monitor systematically their distribution in the pristine steels by means of EBSD. A high density of annealing twins may be considered an inherent weakness of coarse-grained steels, since annealing twins traversing large grains offer paths of easy LBE penetration deep into the steel, depending on the actual grain size and the fraction of twinned grains. As far as dissolution 'pitting' is concerned, the orientation of large grains with annealing twins relative to the steel specimen surface is of primary importance, as shown by Lambrinou et al. [8]. The EBSD orientation maps of Fig. 6 show that all coarse-grained steels (1.4970-r-SA, 316L-p-SA, 316L-r-CW, 1.4970-b-SA, and 1.4970-b-20CW) are characterized by a non-negligible annealing twin density. As indicated by the systematic analysis of EBSD data, the fraction of annealing twins in these steels varied in the 13-60% range; the steels that showed the 'heaviest' combination of grain coarseness with a high fraction of annealing twins were the 1.4970-r-SA (grain diameter >270 µm, 44% twins) and 316L-r-CW (grain diameter >188 µm, 39% twins). The reference MYRRHA fuel cladding 1.4970-t-24CW shows a non-negligible amount of annealing twins (5%), which are smaller due to the fine steel grain size (11-12 µm). Despite their small size, the homogenously distributed annealing twins in this steel are expected to contribute to the local progress of LBE dissolution attack, especially for dissolution depths greater that the width (10-20 μm) of the outer zone that seems to be devoid of such twins.

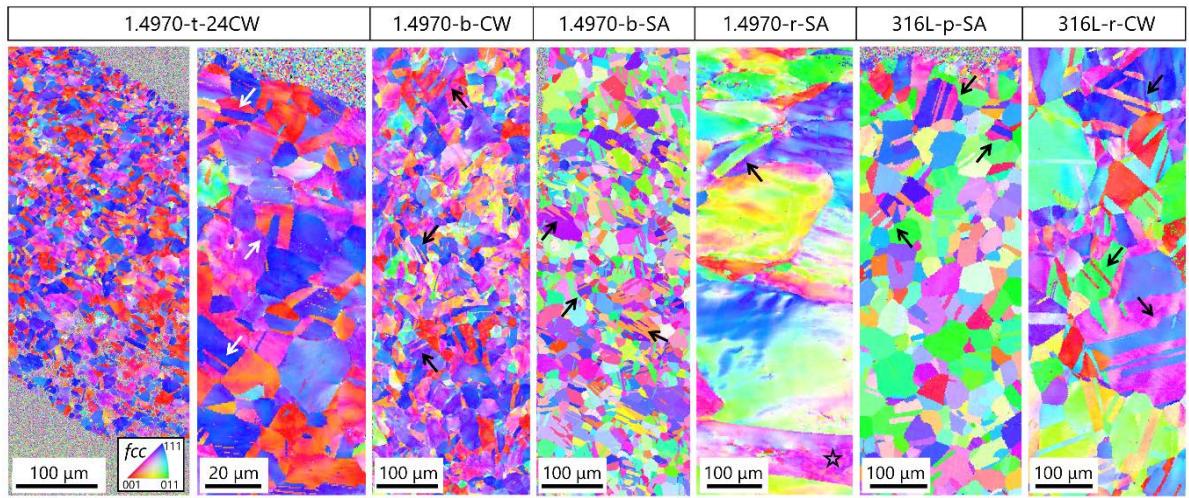

**Fig. 6.** EBSD orientation maps showing annealing twins (arrows) in the 1.4970-t-24CW, 1.4970-b-CW, 1.4970-b-SA, 1.4970-r-SA, 316L-p-SA, and 316L-r-CW steels. The twins are bigger in size in the coarser-grained steels 1.4970-r-SA, 316L-r-CW, and 316L-p-SA. A large elongated grain (star), similar to the ones shown in Fig. 5g, may be observed in 1.4970-r-SA. The inverse pole figure colour maps correspond to the [001] zone axis in all cases, except from those of 1.4970-r-SA and 316L-r-CW that correspond to the [100] and [010] zone axis, respectively.

Another steel microstructural feature that has been clearly shown in earlier studies [8,9,13,47] to promote the progress of LBE dissolution attack in 316L stainless steels are deformation twins. Their density clearly increases in cold worked steels, but due to their fineness it is difficult to monitor their distribution throughout the steel bulk. As the density of deformation twins is higher in cold worked steels, the susceptibility of cold worked steels to dissolution corrosion is expected to be higher, in agreement with previous studies [8,9,13,47]. TEM characterization of the deformation twins in the 316L-r-CW steel heat has been reported earlier by Lambrinou et al. [8]; in that earlier study, the 316L-r-CW steel heat was named 316LH2. LOM images of areas with high deformation twin density in chemically etched 316L-r-CW are shown in Figs. 7a-7b. TEM data, i.e., a bright field (BF) image and a selected area diffraction pattern (SADP), of deformation twins in the highly cold worked 1.4970-t-46CW fuel cladding steel are shown in Fig. 7c and Fig. 7d, respectively. The SADP (Fig. 7d), which was acquired from a set of twin laths (circled area in Fig. 7c) along the <$\bar{1}$01> orientation, verified that the diffraction spots of the steel matrix and twins are connected by a mirror reflection across the {111} twinning plane, as expected for an *fcc* lattice.

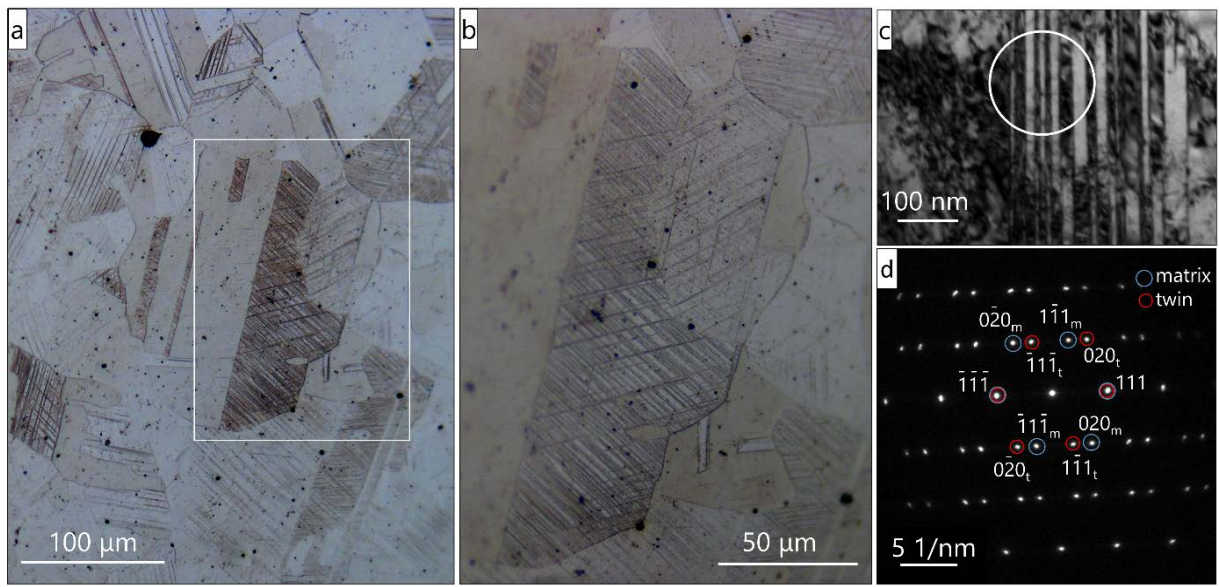

**Fig. 7.** (a,b) LOM images (overview & detail) of fine-scale deformation twins in chemically etched 316L-r-CW. (c) TEM BF image of deformation twin laths in 1.4970-t-46CW, and (d) SADP of the circled area in Fig. 7c.

### 3.2 Quantification of early-stage dissolution corrosion damages

The relative severity of dissolution corrosion damages in the 316L and DIN 1.4970 austenitic stainless steels exposed in this work to oxygen-poor, static liquid LBE at 500°C for 600-1000 h is shown in Fig. 8. Both average (Fig. 8a) and maximum (Fig. 8b) dissolution corrosion depths for all uncoated steels are presented, and the collected data are grouped per test (tests 1-4) so as to allow the direct comparison of dissolution corrosion damages in steels that were exposed to identical LBE conditions, esp. in terms of LBE dissolved oxygen concentration. At a first glance, the depth (average & maximum) of LBE dissolution attack in 1.4970-t-24CW (brushed & polished surface state) is greater for the longer test of 1000 h (test 4) as compared to the shorter tests of 600-619 h (tests 1-3). This is quite straightforward to understand, as the depth of LBE dissolution attack is expected to increase with the time of exposure, for conditions permitting the continued steel dissolution. For steel exposures to baths of static LBE with a finite volume (here, ~0.5 litre), as those performed in this work, steel dissolution is expected to stop (at the absence of dissolved oxygen that can take species out of solution by forming oxides that precipitate out of the bath) when the LBE bath becomes saturated in dissolved species (Ni, Mn, Cr, Fe) at the selected exposure temperature. Taking into account that the solubilities of Ni, Cr and Fe (i.e., the main steel alloying elements) are $2.30 \times 10^{-1}$, $1.60 \times 10^{-3}$ and $2.19 \times 10^{-4}$, respectively, at 500°C [8], saturation of the LBE bath could not have occurred even upon the complete dissolution of the steel specimens in tests 1-4; moreover, complete dissolution of the steel specimens would not have been possible within 600-1000 h, based on the 316L dissolution corrosion rates determined at 500°C in an earlier study [8]. Hence, the herein LMC effects were not inhibited by the size of the test setup, and the reported results are a direct consequence of the LBE exposure conditions ($T$, $C_0$, time) and the exact steel characteristics (composition, microstructure, and thermomechanical state).

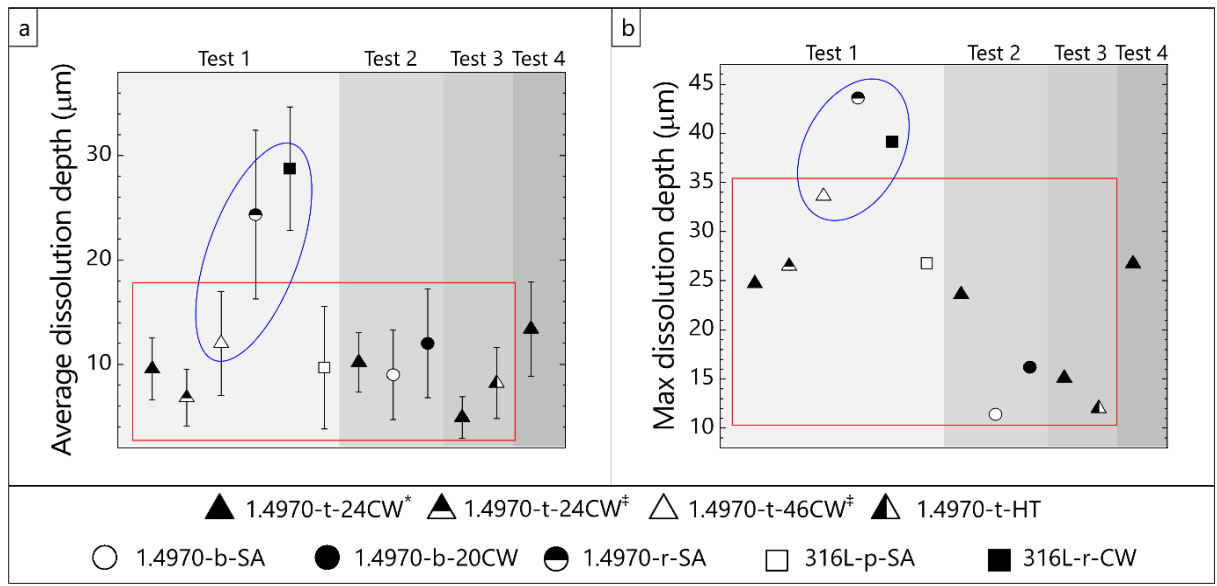

**Fig. 8.** (a) Average and (b) maximum depths of LBE dissolution attack for the 316L and DIN 1.4970 steels exposed in this study to oxygen-poor, static liquid LBE in tests 1-4. The red frames indicate the variation of dissolution depth values in DIN 1.4970 fuel cladding tubes for the first ~600 h of exposure; the blue frames indicate the 3 steels (1.4970-r-SA, 316L-r-CW, 1.4970-t-46CW) that suffered the most severe LBE dissolution attack within the first 619 h of exposure.

When assessing the resistance of a specific steel to dissolution corrosion, the magnitude of the maximum depth is of great interest, as it can give an indication of the time frame within which the integrity of thin-walled components, such as fuel cladding tubes and heat exchanger tubes, can be compromised. The wall thickness of the MYRRHA reference DIN 1.4970 fuel cladding tubes is 450 µm, while a typical wall thickness of heat exchanger tubes made of 316L steels is about 1 mm. As may be seen in the plot of Fig. 8b, the steels that showed the most severe LBE dissolution attack are 1.4970-r-SA (~43.5 µm), 316L-r-CW (~39 µm), and 1.4970-t-46CW (~34 µm), after 619 h of LBE exposure. As discussed in section 3.1, 1.4970-r-SA and 316L-r-CW are characterised by the most unfavourable combination of grain coarseness (>270 µm for 1.4970-r-SA, >188 µm for 316L-r-CW) with high fractions of annealing twins (44% for 1.4970-r-SA, 39% for 316L-r-CW). Hence, it is clear – and will be corroborated by supportive data in sections 3.3 & 3.4– that very coarse-grained stainless steels with high density of annealing twins are very susceptible to dissolution corrosion and their use should be avoided in Gen-IV LFRs. The third most severely affected steel, the 1.4970-t-46CW, was very fine-grained (~5.5 µm) and had a rather low fraction of annealing twins (2%). In this particular steel, however, the density of deformation twins was very high due to the high degree of cold deformation, and the overall GB area was the largest of all steels used in this study. As was shown in prior studies [8,9,13,47], both deformation twins and GBs are preferred paths for the LBE ingress into the steel bulk. Therefore, steels that are heavily cold worked, with high densities of deformation twins and very fine grain sizes are undesirable for use in Gen-IV LFRs, due to their enhanced susceptibility to dissolution corrosion.

Going forth with the comparison of steel performance, it is clear that an enhanced degree of cold work in 'similar' steels exposed to identical test conditions increases the maximum depth of LBE dissolution attack. This becomes clear by comparing 1.4970-t-24CW (~24.5 μm) with 1.4970-t-46CW (~34 μm) (test 1); 1.4970-b-SA (~11.5 μm) with 1.4970-b-20CW (~16 μm) (test 2); and 316L-p-SA (~27 μm) with 316L-r-CW (~39 μm) (test 1). Typically, a higher degree of cold work means higher density of deformation twins and grain refinement (i.e., larger GB area). This is true for the 1.4970-t-24CW/1.4970-t-46CW pair in this work, thus explaining the more severe attack in 1.4970-t-46CW as compared to 1.4970-t-24CW, but it is not exactly true for the other two steel pairs. As mentioned in section 3.1, 1.4970-b-SA was finer-grained (18 μm) than 1.4970-b-20CW (30 μm) due to the steel processing route; moreover, 1.4970-b-SA had only 13% annealing twins, while 1.4970-b-20CW had an extreme annealing twins fraction of 60%. Therefore, as far as this pair of steels is concerned, the deeper LBE dissolution attack observed in 1.4970-b-20CW may be attributed to the combination of coarser grains and numerous annealing twins (on the micrometre scale) with a higher density of deformation twins (on the nanometre scale). Similarly, the much deeper dissolution attack in 316L-r-CW as compared to 316L-p-SA can be associated with the fact that 316L-r-CW was coarser-grained (68-188 μm vs. 44-51 μm), had a higher fraction of annealing twins (39% vs. 21%), and an expected higher density of deformation twins due to the heavier cold working treatment that has been confirmed on the nanometre scale by TEM analysis of the two steels in an earlier study [8].

Mild differences in surface roughness do not appear to affect the extent of dissolution damages in DIN 1.4970 fuel cladding tubes. The maximum corrosion depth in brushed & polished 1.4970-t-24CW (~24.5 μm) is slightly smaller than the corresponding depth in brushed 1.4970-t-24CW (~26.5 μm), which can be tentatively attributed to the rougher initial surface of brushed 1.4970-t-24CW ($R_{max} \approx 3.91$ μm) as compared to brushed & polished 1.4970-t-24CW ($R_{max} \approx 3.65$ μm). Taking into account the small differences in surface roughness of the two fuel cladding tubes, as well as the short steel exposure (619 h), it is difficult to assess the effect of steel surface roughness on its resistance to dissolution corrosion. One should not completely disregard this aspect of steel preparation for Gen-IV LFRs, though, especially when it comes to structural components that are exposed to relatively high temperatures (e.g., heat exchangers, core support plate, core barrel in the MYRRHA system), as in these parts of the reactor system, surface roughness specifications are more relaxed (as opposed to the required surface state of fuel cladding tubes), and undesirable dissolution corrosion effects might be more severe due to the high temperatures. Based on the collected data in this work, it might be safe to suggest that rougher steel surfaces result in deeper dissolution, as surface defects (e.g., skin folds and other imperfections) form an inhomogeneous diffusion boundary layer at the steel/LBE interface, which might either create localised crevice corrosion conditions (e.g., inside a skin fold) under static LBE conditions or inhibit the satisfactory refreshment of oxygen supply inside grooves, scratches and other surface undulations under flowing LBE conditions.

The effect of texture on the dissolution corrosion behaviour of the steels studied in this work can be assessed only indirectly based on the collected corrosion data. As mentioned in section 3.1, grains with the <111> orientation that lie parallel to the direction of tensile load applied during steel cold deformation have a higher density of deformation twins, as also confirmed in a study of 316L steel dissolution corrosion by Klok et al. [9]. These grains are inherently more susceptible to LBE dissolution attack due to the higher density of deformation twins, which promote the faster ingress of LBE into the steel bulk. In this work, the stainless steels with strong <111> texture were the 1.4970-t-46CW, 1.4970-t-24CW, 1.4970-b-20CW, and 316L-r-CW (Fig. 1). On the other hand, the solution-annealed steels 1.4970-r-SA, 1.4970-b-SA, 1.4970-t-HT, and 316L-p-SA show a much more random crystal orientation, and a couple of them (1.4970-r-SA and 316L-p-SA to a lesser degree) shown an almost complete absence of the <111> orientation (Fig. 1). Considering the collected corrosion data, one might point out the following: (i) not all steels with a strong <111> texture component show deep corrosion damages (e.g., the maximum depth of dissolution attack in 1.4970-b-20CW is only ~16 µm), and (ii) not all steels with a weak <111> texture component exhibit better resistance to dissolution corrosion (e.g., the maximum depth of dissolution attack in 1.4970-r-SA is 43.5 µm). Based on the above, one might suggest that texture is not the decisive factor when manufacturing steels suitable for the construction of Gen-IV LFRs. This and prior studies [9,13] have directly or indirectly touched upon the importance of steel texture; however, even though grain orientation is expected to play a role in the local advancement of the dissolution front [9], it does not define the overall resistance of a steel to dissolution corrosion.

High-temperature annealing on the 'reference' 1.4970-t-24CW fuel cladding had a mild effect on its resistance to dissolution corrosion. As mentioned in section 3.1, annealing (1000°C, 2 h) of the 1.4970-t-24CW resulted in grain coarsening (from 12 µm to 27 µm), a steep increase in the fraction of annealing twins (from 5% to 47%), and a change of the preferred grain orientation from a mixed <111> and <001> texture to a more random one (Fig. 1). Despite the sharp increase in the fraction of annealing twins, the intermediate grain size combined with the (expected) reduction in deformation twins accounts for the more shallow maximum dissolution depth in 1.4970-t-HT when compared to 1.4970-t-24CW (Fig. 8b). Studying the effects of high-temperature annealing on the LMC behaviour of fuel cladding steels, such as the DIN 1.4970 steel used in this work, is of interest for the deployment of Gen-IV LFRs, as high-temperature transients cannot be excluded during the reactor operation. This work is a first step towards the understanding of such high-temperature excursions on the (fuel cladding) steel dissolution corrosion behaviour.

Before concluding the discussion on the relative performance of the steels used in this work, based on the data shown in Fig. 8, one might want to address the effect of steel composition on the relative resistance to dissolution corrosion. The concentrations of the highly soluble in LBE steel alloying elements Ni, Cr and Mn (in mass%) were: (i) 15.05, 15.06 and 1.86, respectively, for the DIN 1.4970 steel heat, and (ii) 9.97-10.10, 16.68-16.73, and 1.75-1.81, respectively, for the two 316L steel heats. Even though one might expect higher dissolution corrosion rates for the steel with higher concentration in Ni, i.e., the DIN 1.4970, due to the fact that Ni is the steel alloying element with the highest solubility in LBE, the LBE-exposed fuel cladding tubes showed maximum dissolution corrosion damages that were comparable to those measured in 316L-p-SA. Moreover, the depth of dissolution corrosion damages in the DIN 1.4970 fuel cladding tubes were rather moderate, when compared to all other steels tested in this work (see Fig. 8a). On the same topic, a prior study by Klok et al. [47] reported lower local LBE dissolution attack in Ni-rich bands found in a 60% cold-deformed specimen of the 316L-p-SA steel heat that is also used in this work. This counter-intuitive observation was attributed to the effect of Ni on the steel stacking fault energy (SFE): Ni increases the material's SFE, limiting the formation of defects, such as deformation twins, during plastic deformation. The Ni-rich bands in 60% cold-worked 316L-p-SA were indeed characterised by a lower density of deformation twins, which resulted in the locally slower advancement of the dissolution front in these bands during the steel exposure to oxygen-poor, static LBE at 500°C for 1000 h [47]. It is, thus, reasonable to assume that the compositional differences of the 316L and DIN 1.4970 steels used in this work on the magnitude (depth) of dissolution corrosion damages could have been compensated by steel microstructural features affecting the steel dissolution corrosion (grain size, density of annealing/deformation twins, etc.). It should be pointed out that even though the 1.4970-r-SA (richer in Ni than 316L) is characterised by a very low density of deformation twins, it suffers the highest dissolution corrosion damages (Fig. 8b), primarily due to other microstructural features promoting dissolution corrosion (i.e., coarse grains, large fraction of annealing twins). The second most-affected steel, i.e., the 316L-r-CW, has a similar microstructure in terms of grain coarseness and fraction of annealing twins, plus a much higher density of deformation twins; however, the maximum depth of LBE dissolution attack in 316L-r-CW is ~10% lower than that in 1.4970-r-SA exposed to identical conditions (test 1). In the case of this specific steel pair, one could postulate that steel compositional differences (esp. the higher Ni content of 1.4970-r-SA compared to 316L-r-CW) might indeed be accountable for the observed differences in corrosion behaviour.

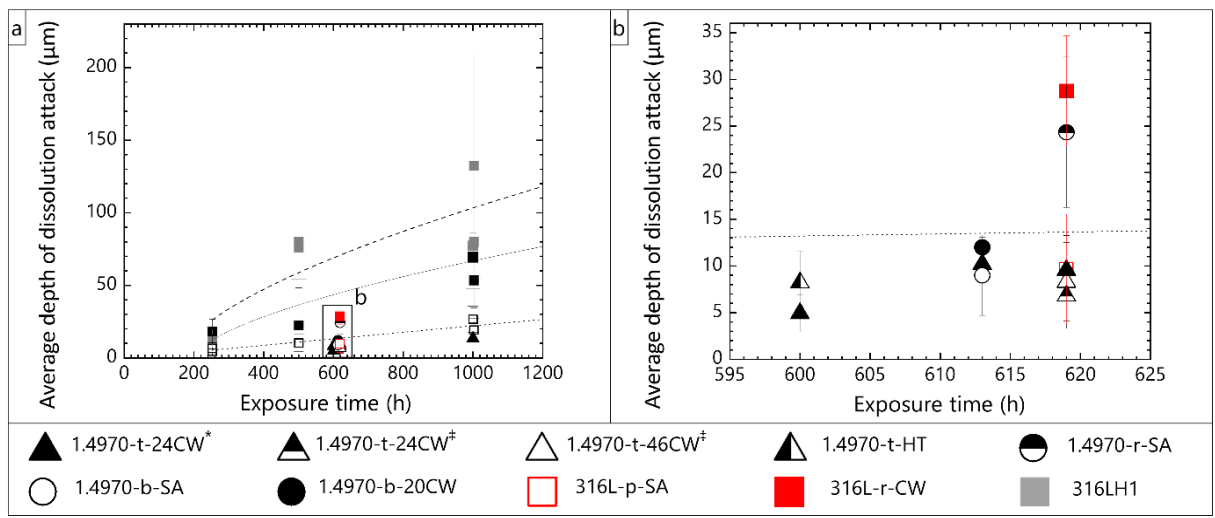

**Fig. 9.** (a,b) Average depth of LBE dissolution attack in the 316L and DIN 1.4970 steels exposed to oxygen-poor liquid LBE at 500°C for ~600 h in this work: the acquired data are enveloped by a frame in Fig. 9a and are plotted in Fig. 9b. The acquired data are plotted with average dissolution attack data reported earlier [8] for 316L steels (316L-p-SA, 316L-r-CW, 316LH1) exposed to oxygen-poor LBE at 500°C for 253-1003 h. The 316L-p-SA and 316L-r-CW heats used in this work were earlier named 316LSA and 316LH2, respectively [8].

The (average & maximum) LMC data collected in this work are put in perspective with respect to 316L dissolution corrosion data acquired in a previous study [8]. In that study, Lambrinou et al. [8] exposed simultaneously to oxygen-poor, static LBE at 500°C for 253-3282 h three different 316L steel heats, i.e., one solution-annealed heat (316LSA; this heat is here named 316L-p-SA) and two cold-worked heats (316LH1 & 316LH2; the 316LH2 heat is here named 316L-r-CW) [8]. The early-stage (600-1000 h) dissolution corrosion data acquired in this work from both 316L and DIN 1.4970 steels are plotted as function of time in Figs. 9-10 together with the early-stage (253-1003 h) 316L dissolution corrosion data reported in Ref. [8]. The 316L data acquired in this work on two of the steel heats used in Ref. [8] add new data points (619 h; test 1) from exposing these 316L steel heats to comparable test conditions (autoclaves of identical design, same conditioning gas). One may observe in the graphs in Figs. 9-10 that, even though the depths of 316L dissolution corrosion damages measured in this work are on the lower end of the dissolution depth ranges reported in Ref. [8], they still fall within these ranges. It is difficult to attribute the (consistently for both steel heats) relatively shallower depths of LBE dissolution attack in the same 316L steels to fluctuations in the LBE oxygen concentration (Figs. 2-3) due to the imperfect leak tightness of the used autoclaves, as similar fluctuations have also been reported in Ref. [8]. An unobtrusive difference in the supply of the conditioning gas might explain these small differences in the severity of the dissolution corrosion effects observed in this work for the same two 316L steel heats as compared to Ref. [8]: in the earlier tests [8], the reducing gas (Ar-5% $H_2$) was supplied close to the bottom of the LBE bath by means of a 'bubbling tube', thus ensuring a form of mild

LBE mixing during steel exposure. In this work, the gas 'bubbling tube' has been removed, and the reducing conditioning gas was supplied directly to the gas plenum, making all LMC processes strictly diffusion-controlled. The complete lack of LBE mixing establishes unavoidably thicker diffusion boundary layers at the steel/LBE interfaces, slightly slowing down the steel dissolution process. Moreover, it is worthwhile mentioning that there is a non-negligible difference between Ref. [8] and this work with respect to the overall steel specimen surface that was exposed to the same LBE volume (~0.5 litre): the overall steel specimen surface in Ref. [8] was 911 mm² for all steel exposures in the 253-1033 h time frame, while in this work the same surface varied in the 3490-3830 mm² range for tests 1-3 (this calculation considers both inner and outer tube surfaces, as the tubes were open during testing). The combined lack of LBE mixing and the significantly larger steel surfaces exposed to the same LBE volume account for the lower dissolution corrosion depths measured in this work as compared to Ref. [8]. It is quite reasonable to assume that the concentration of dissolved steel alloying elements in the thicker (due to the complete lack of LBE mixing) diffusion boundary layers at the steel/LBE interfaces would quickly reach levels (due to the larger steel surfaces) that could delay the progress of dissolution corrosion, as the diffusion-based removal of dissolved species from these layers is slower than in the presence of mild mixing.

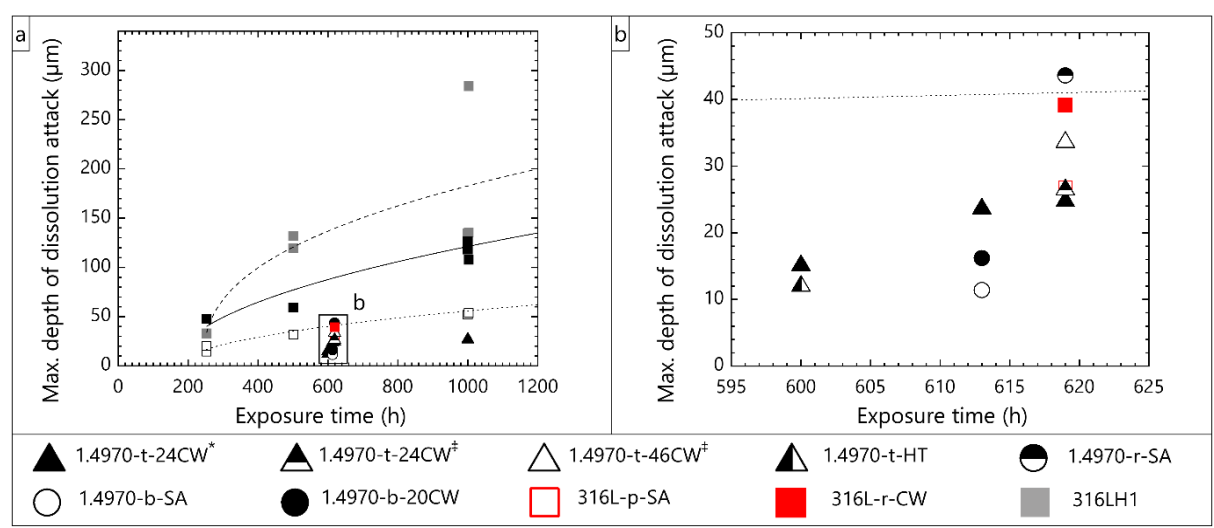

**Fig. 10.** (a,b) Maximum depth of LBE dissolution attack in the 316L and DIN 1.4970 steels exposed to oxygen-poor liquid LBE at 500°C for ~600 h in this work: the acquired data are enveloped by a frame in Fig. 10a and are plotted in Fig. 10b. The acquired data are plotted with maximum dissolution attack data reported earlier [8] for 316L steels (316L-p-SA, 316L-r-CW, 316LH1) exposed to oxygen-poor LBE at 500°C for 253-1003 h.

The newly acquired 316L dissolution corrosion data on two of the steel heats previously used [8] are used to 'anchor' the set of data collected from all steels in this work, providing a perspective on the relative resistance of DIN 1.4970 steels to dissolution corrosion in view of the lack of prior published data. As shown in Fig. 8, the dissolution corrosion damages in as-fabricated DIN 1.4970 fuel cladding tubes (1.4970-t-24CW, 1.4970-t-46CW) are roughly comparable with such damages in 316L-r-SA and 316L-r-CW (the latter benchmarking the "worst case scenario" in terms of LMC resistance). Therefore, one might safely postulate that the seemingly underestimated severity of corrosion damages in the 316L steel heats exposed in this work (by ~25% and ~50% for average and maximum corrosion depths, respectively) are likely to translate to the DIN 1.4970 steels that were exposed to similar conditions in tests 1-4. This puts forth the necessity of performing future tests on DIN 1.4970 stainless steels that ensure some form of mild LBE mixing during exposure, so as to assess the true resistance of such steels to dissolution corrosion (these tests cannot be performed in a forced convection loop due to the forbiddingly low LBE oxygen levels that would surely damage the steels used for the loop construction, esp. for long-term testing).

*3.3    Early-stage dissolution corrosion damages: overview*

The quantification of dissolution corrosion damages allows the (conditional) assessment of the overall steel resistance to LBE dissolution attack; this assessment is considered 'conditional', as it is unavoidably affected by the test setup limitations (e.g., LBE bath size, presence/absence of LBE mixing), the testing approach (e.g., LBE preconditioning approach prior to steel immersion, exposure of several steel heats with distinctly different microstructures and thermomechanical states), the success in controlling and monitoring the exposure conditions (e.g., by using oxygen sensors, active oxygen control systems), etc. (see also section 3.2). However, full appreciation of all aspects of the dissolution corrosion behaviour of a specific steel can only be achieved via the detailed microstructural analysis of the exposed steel specimens. Such analysis is needed so as to shed light on the steel microstructural features affecting the dissolution corrosion mechanism on the micrometre and nanometre scales. Typical examples of LBE dissolution attack in all 316L and DIN 1.4970 stainless steels exposed in this work to oxygen-poor, static LBE at 500°C for ~600 h are shown in Fig. 11. These examples are SEM images taken with a backscattered electron (BSE) detector, in order to facilitate the inspection of the LBE-affected zones due to their compositional contrast with the unaffected steel bulk. In all cases, both overview and detailed BSE images are provided, so as to capture the dissolution corrosion damages in two different scales. The images of DIN 1.4970 fuel cladding tubes (1.4970-t-24CW, 1.4970-t-46CW, 1.4970-t-HT) in Fig. 11 are taken from the outer tube surfaces, as these surfaces are of primary importance when assessing the compatibility of a fuel cladding steel with a HLM coolant; both outer and inner tube surfaces suffered LBE dissolution attack, though, as the tubes were immersed open in the LBE bath.

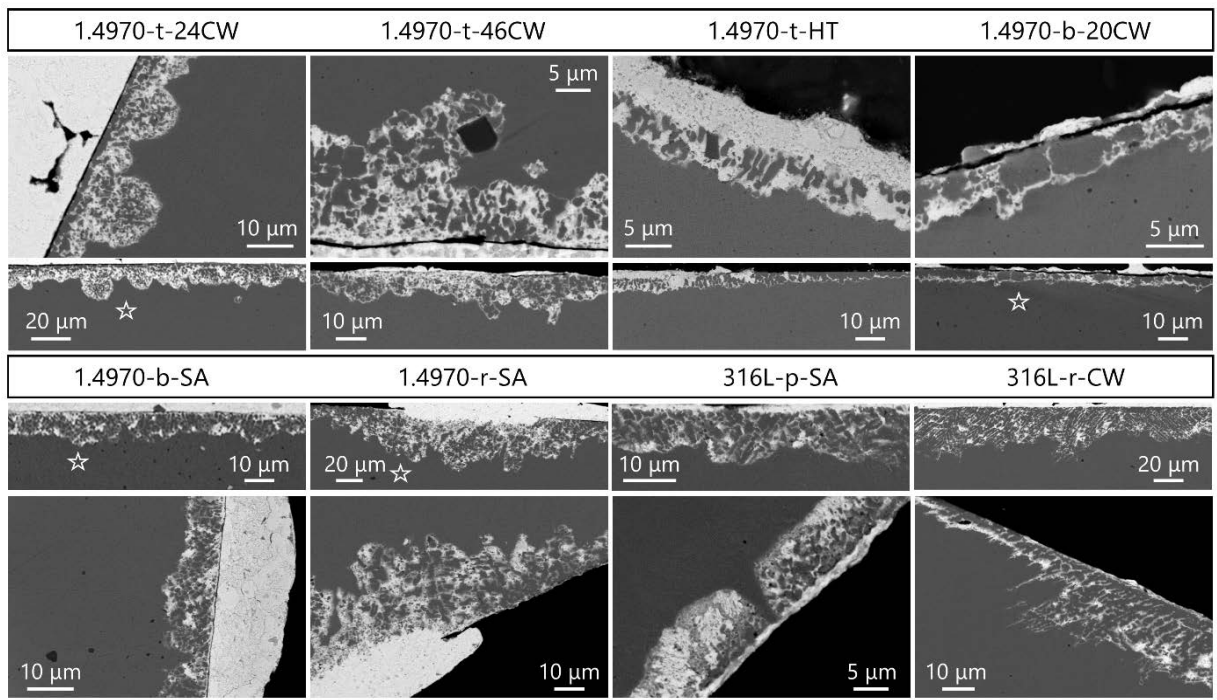

**Fig. 11.** BSE images (overviews and details) of the typical appearance of LBE dissolution attack in the 316L and DIN 1.4970 steels exposed to LBE for ~600 h in this work. Stars indicate locations in the overviews that appear magnified in the details. BSE images showing dissolution corrosion effects in 1.4970-t-24CW, 1.4970-t-46CW, 1.4970-r-SA, 316L-p-SA, and 316L-r-CW from test 1; 1.4970-b-SA, and 1.4970-b-20CW from test 2; 1.4970-t-HT from test 3.

In general, LBE dissolution attack in all 316L and DIN 1.4970 stainless steels used in this work started intergranularly, as GBs are preferred paths for the LBE ingress into the steel bulk [8]. The intergranular LBE penetration is typically followed by selective leaching of steel alloying elements (Ni, Mn, Cr), which are transported away by diffusion through the three-dimensional network of LBE penetrations that connect the steel interior with the LBE bath [8]. The selective removal of the austenite stabilisers Ni and Mn leads to the ferritisation of the dissolution-affected zone, a LMC effect that has been understood for 316L steels [8,19], and will be addressed for the DIN 1.4970 steel in section 3.5. Other preferred paths of LBE penetration into the steel bulk, especially cold-worked steels, are deformation twin boundaries, as previously reported for 316L steels [8,9,13,47]. The involvement of deformation twin boundaries in the dissolution corrosion process result in more perturbed dissolution fronts in cold-worked steels (e.g., 316L-r-CW, Fig. 11) when compared to solution-annealed steels (e.g., 316L-p-SA, Fig. 11). Intergranular LBE penetration and selective leaching-induced ferritisation are followed by the transgranular consumption of the ferritized steel grains in the dissolution-affected zone (e.g., 1.4970-t-HT, 1.4970-r-SA, 316L-p-SA, Fig. 11). Sites of locally-enhanced LBE dissolution attack (i.e., dissolution 'pits') are primarily observed in coarse-grained steels with a high fraction of annealing twins, such as 1.4970-r-SA (see Fig. 11) and 316L-r-CW. Lambrinou et al. [8] have previously reported the formation of a big (~100 μm in depth, ~370 μm in width) dissolution 'pit' on the surface of a 316L-r-CW specimen exposed to oxygen-poor ($C_0 < 2\times10^{-8}$ mass%) static liquid LBE at 500°C for 1000 h; that 'pit' formed on the steel surface due to the inward convergence of annealing twins in large (>165 μm) sub-surface grains. Overall, the dissolution corrosion process does not proceed perpendicular to the steel surface, but changes locally direction following preferred paths in the steel microstructure (e.g., GBs, annealing/deformation twins). This clear interplay of the dissolution corrosion process with the steel microstructure accentuates the importance of the latter, as this is very often accountable for highly undesirable LMC effects, such as dissolution 'pitting'. Hence, as pointed out in this and prior studies [8,9,13,47], the optimisation of the steel microstructure, in terms of grain size distribution and density of dissolution corrosion-promoting defects such as annealing/deformation twins, is of paramount important before using a specific steel grade in the construction of Gen-IV LFRs.

BSE images of the time evolution of dissolution corrosion damages in the 316L and DIN 1.4970 steel tested in this work are shown in Figs. 12 and 13, respectively. Fig. 12 includes unpublished images of dissolution corrosion damages in 316L-p-SA and 316L-r-CW exposed to oxygen-poor ($C_0 < 10^{-8}$ mass%), static liquid LBE at 500°C for 253, 501 and ~1000 h in an earlier study [8]. In this work, the same 316L steel heats were exposed oxygen-poor ($C_0 < 10^{-9}$ mass%), static LBE at 500°C for 619 h (test 1). The early stages (253-1000 h) of LBE dissolution attack the industrial size solution-annealed 316L-p-SA show that attack starts intergranularly (Fig. 12a), ferritising the first layer of grains (2-4 μm) underneath the specimen surface. As time progresses and the LMC damages deepen (Figs. 12c and 12e), LBE dissolution attack becomes transgranular in the affected zone, gradually consuming the Fe-based (ferritized) grains until the whole dissolution zone loses any resemblance with the initial steel microstructure. In a real nuclear system and in the presence of LBE flow, the whole dissolution-affected zone will be completely removed, as it essentially filled by LBE with high concentrations of dissolved steel alloying elements; in completely static tests, as those performed in this work, the damaged zone is retained and may be studied by SEM, revealing the loading of LBE with dissolved species (darker, featureless areas in the damaged zone, Fig. 12e).

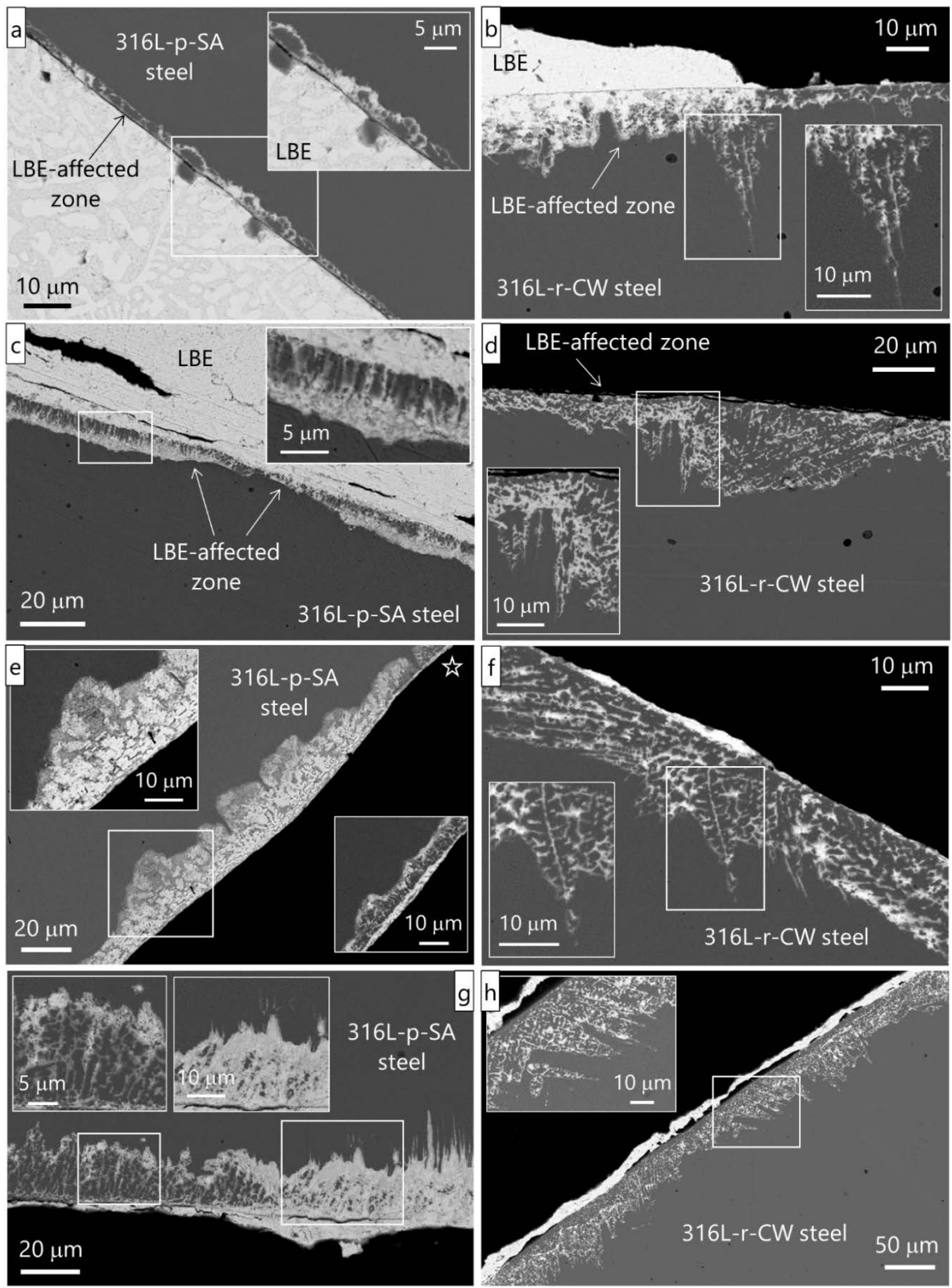

**Fig. 12.** BSE images of the time evolution of LBE dissolution attack in the 316L-p-SA and 316L-r-CW steels (316LSA and 316LH2, respectively, in [8]) exposed to oxygen-poor, static LBE at 500°C for (a,b) 253, (c,d) 501 and (g,h) ~1000 h (previous work [8], unpublished data) and (e,f) 619 h (this work).

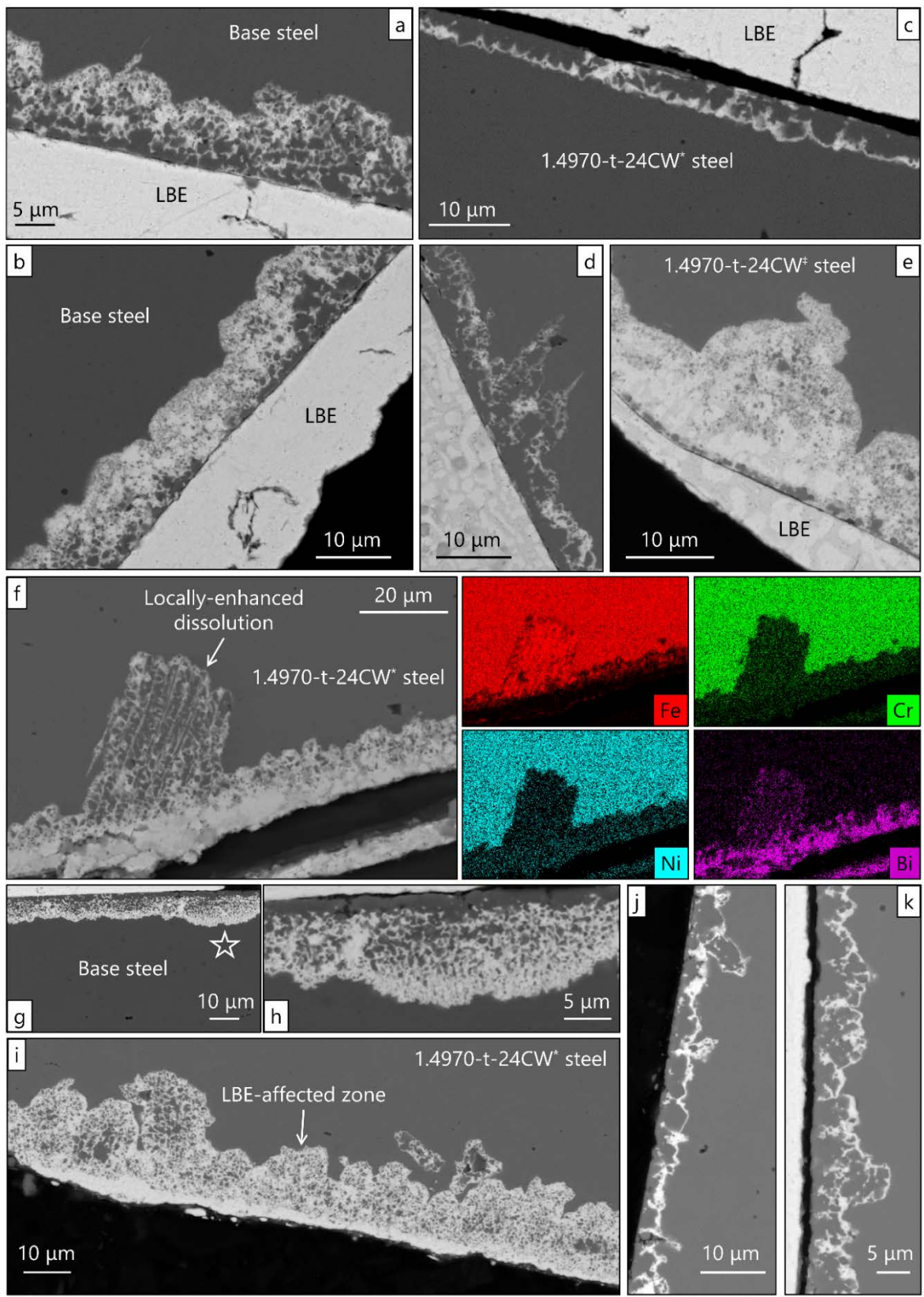

**Fig. 13.** BSE images of the time evolution of LBE dissolution attack in 1.4970-t-24CW exposed to oxygen-poor, static LBE at 500°C for (a-f) ∼600 h and (g-k) 1000 h. (a-c) Variable depths (4-12.5 μm) of attack in 1.4970-t-24CW (brushed & polished). Dissolution attack starts intergranularly (c) and proceeds transgranularly (a,b) with depth. (d,e) Dissolution front perturbations in 1.4970-t-24CW (brushed). (f) Site of locally-enhanced LBE dissolution attack (>40 μm) in 1.4970-t-24CW (brushed & polished). (g-k) Variable depths (4-26 μm) of attack in 1.4970-t-24CW (brushed & polished).

As discussed in section 3.2, in primarily static LBE tests, the diffusion barrier layers that are established at the steel/LBE interfaces become at a certain point too thick, thereby slowing down the progress of dissolution corrosion, as the dissolved in LBE steel alloying elements must diffuse through these layers and into the main volume of the LBE bath for steel dissolution to continue at a pace dictated by the exposure temperature (i.e., when dissolution is driven by the elemental chemical potential differences between solid steel and 'fresh', non-contaminated liquid LBE at that temperature). For longer exposures, when the thickness of the dissolution-affected zone has also increased a lot, the dissolved in LBE penetrations steel alloying elements must diffuse outwards through the thick LBE-affected zone. This results in a deceleration of the steel dissolution when the thickness of the LBE-affected zone exceeds a (temperature-dependent) value, as reported by Lambrinou et al. [8] in an earlier study on the dissolution corrosion behaviour of 316L steels. In the early stages of dissolution corrosion, though, when the LBE-affected zone is still relatively thin, the dissolution process is expected to progress more or less in a uniform manner into the steel bulk. However, this is not what is observed in 316L-p-SA exposed for 619 h to oxygen-poor, static LBE in this work (Fig. 12c), where the LBE dissolution does not continue inwards intergranularly, but changes into pure transgranular close to the dissolution front. A closer inspection of the BSE image of Fig. 12c reveals that the dissolution-affected zone is completely featureless close to the dissolution front, suggesting the complete consumption of the steel in that area. One possible way to explain this is by considering that, under truly static conditions, the thick diffusion boundary layer at the steel/LBE interface delays the outward diffusion of steel alloying elements into the main LBE bath volume, putting an obstacle to the deepening of attack. Under these conditions, the LBE that has reached a certain depth into the steel remains for a prolonged period of time at that depth, completely dissolving the steel at the dissolution front, while the surface layer of ferritized grains is still present. Let us not disregard the fact that the deeper the LBE penetration, the more depleted it becomes in oxygen, hence, its effectiveness in dissolving the steel is locally enhanced.

The change in dissolution corrosion mode from primarily intergranular (left inset, Fig. 12g) to primarily transgranular (right inset, Fig. 12g) over a short distance (~70 μm) on the steel surface may be attributed to compositional inhomogeneities in industrial-size stainless steel heats, such as the 316L-p-SA (cold rolled, 15 mm-thick plate) used in this work. The effect, known as 'chemical banding', has been observed in 60% cold-deformed 316L-p-SA exposed to oxygen-poor, static LBE at 500°C for 1000 h by Klok et al. [47]. The effect of local steel composition on the transition from selective to non-selective leaching in the 316L-p-SA steel was touched upon earlier by Lambrinou et al. [8], who identified the possible association of the local steel chemical composition (esp. its Ni content) with the local density of deformation-induced defects that promote dissolution attack.

Dissolution corrosion in cold-worked stainless steels, such as the 316L-r-CW used in this work, progresses by using preferred paths of LBE ingress (GBs, deformation twin boundaries) into the steel bulk. The higher density of deformation twins in such steels results in dissolution fronts that are highly perturbed; sharp protuberances of the dissolution front (Figs. 12b, 12d, 12f, 12h) are expected to be characterised by high stress intensities in their immediate vicinity, which may nucleate cracks in radiation-hardened stainless steels, as suggested earlier [8]. One should not neglect the fact that the reduced ductility of austenitic stainless steels that have undergone in-service radiation embrittlement (typically saturating at ~10 dpa in stainless steels [49-51]) is associated with a reduced resistance to crack propagation; therefore, one should try to minimise the density of stress concentrators (e.g., sharp perturbations of the dissolution front) that can nucleate cracks in such steels. This is particularly important for thin-walled components, such as fuel cladding tubes and heat exchanger tubes, as easier crack propagation in radiation-hardened stainless steels might lead to the premature in-service loss of the component structural integrity. Notwithstanding the need for producing stainless steels without deformation-induced defects, such as deformation twins, for a good resistance to dissolution corrosion, one must seek viable compromises between the possible problems created by LMC and meeting other steel property requirements that demand a degree of cold work (e.g., resistance to radiation & thermal creep).

Typical BSE images of early-stage (600-1000 h) dissolution corrosion effects in the reference MYRRHA fuel cladding material 1.4970-t-24CW with different surface states (brushed & polished vs. brushed) are shown in Fig. 13. Similar to the 316L steels used in this work, dissolution attack starts intergranularly (Fig. 13c) and proceeds transgranularly (Fig. 13a-13b). The occurrence of both intergranular and transgranular dissolution modes on the same cross-section of 1.4970-t-24CW, after only a short (~600 h) exposure to liquid LBE, cannot be attributed to compositional inhomogeneities, as this particular steel is very homogeneous, while its high Ni content (15.05%) is expected to increase the steel SFE to levels that maintain the density of dissolution-promoting deformation twins within reasonable limits (i.e., limits that do not promote the complete steel dissolution within very short exposure periods). It can be assumed, therefore, that the truly static LBE exposure conditions imposed in this work might be (at least partly) accountable for the intergranular to transgranular transition in dissolution mode, due to the establishment of a thick diffusion boundary layer at the steel/LBE interface. This transition in dissolution mode becomes more pronounced in the longer (1000 h) exposure of 1.4970-t-24CW (Figs. 13g, 13h, 13i), which further supports the idea that this transition is associated with the fact that the outward diffusion of dissolved steel alloying elements is obstructed upon testing in the complete absence of LBE mixing by thicker diffusion boundary layers at the steel/LBE interface. The limited pool of 1.4970-t-24CW corrosion data acquired in this study, however, does not permit any firm conclusions on this aspect. The simultaneous occurrence of intergranular and transgranular dissolution modes is not affected by the surface state of the pristine 1.4970-t-24CW, as it was also observed on the same cross-section of brushed 1.4970-t-24CW (Figs. 13d-13e).

Even though the formation of very deep dissolution 'pits' did not characterise the pristine fuel cladding tubes 1.4970-t-24CW and 1.4970-t-46CW, presumably due to the grain fineness of these steels, a BSE image of a quite deep (~42 μm) 'pit' may be observed in Fig. 13f together with the corresponding EDS maps that show selective leaching of Ni and Cr. This is the only case of severe dissolution 'pitting' observed in 1.4970-t-24CW in this work (test 2), therefore, this data point was omitted from the plot of Fig. 8b, being considered an aberration of the material's behaviour. The cause of its formation is unclear, and it could likely be attributed to a local steel inhomogeneity that is atypical for the material and has formed during steel production. The occurrence of such 'pits' is highly undesirable for thin-walled components, such as fuel cladding tubes; hence, their origin must be understood and ideally eliminated by means of controlled steel manufacture.

The depth of LBE dissolution attack varied on the 1.4970-t-24CW fuel cladding tube surface from a few (3-6 μm) micrometres just below the outer steel surface (Figs. 13c, 13j, 13k) to many micrometres in the more affected areas (Figs. 13a, 13b, 13g, 13h, 13i). There is no obvious cause for these variations in dissolution depth, other than the failure of a pre-existent oxide scale, which was either a native oxide that formed in air prior to the steel immersion in the LBE bath or a thin oxide film that formed upon steel immersion in the LBE bath (typically this was accompanied by a further decrease in the already low LBE oxygen concentration, $C_O < 10^{-8}$ mass%, at the moment of steel immersion, as reported earlier [8]). Irrespective of the nature of the initial oxide scale on the fuel cladding steel surface, its prolonged exposure to oxygen-poor LBE led to its deterioration, allowing the onset of LBE dissolution attack in these areas.

### 3.4  *Steel microstructural features promoting dissolution corrosion*

### 3.4.1  *Grain boundaries & annealing/deformation twins*

Grain boundaries (GBs) are preferred paths for the LBE ingress into the steel bulk; this has been established in an earlier study on the dissolution corrosion behaviour of 316L steels at 500°C [8], and also in this work, as it has been shown that in the tested 316L and DIN 1.4970 steels, LBE dissolution attack starts invariably intergranularly (Figs. 11-13). As discussed in section 3.2, very coarse-grained steels with high fractions of annealing twins (e.g., 1.4970-r-SA, 316L-r-CW) as well as very fine-grained steels with high fractions of deformation twins (e.g., 1.4970-t-46CW) are very susceptible to dissolution corrosion. This essentially summarises the effect of the steel grain size distribution on the steel resistance to dissolution corrosion and should not be discussed further. The steel grain size must be clearly optimised together with the steel-specific deformation process and the deformation-induced defect population to meet the property requirements of the targeted reactor component, while minimising the steel susceptibility to LMC effects to the extent possible. When this is not achievable with conventional steel metallurgy, surface engineering technologies, such as anticorrosion coatings (see section 3.6) and surface modification, and/or development of new (alumina-forming) stainless steels must be considered.

Annealing and deformation twins are very potent steel defects when it comes to promoting dissolution corrosion in stainless steels; this has been firmly established in previous studies on the dissolution corrosion of 316L stainless steels exposed to oxygen-poor, static LBE at 500°C [8,9,13,47]. Additional evidence of the fact that annealing and deformation twins constitute preferred paths for the LBE ingress into 316L steels is provided in Figs. 14a-14b and 14c-14d, respectively. These LOM images were acquired from OPS-polished specimens of the 316L-r-CW steel, which showed one of the three most severe dissolution corrosion damages in this study (Fig. 8b). The particular steel specimen shown in Fig. 14 (unpublished data) was exposed to oxygen-poor static LBE at 500°C for 1003 h in the framework of a prior study [8]; however, the observed contribution of annealing/deformation twins to the dissolution process is very important for the in-depth understanding of the early-stage (<1000 h) dissolution corrosion behaviour of stainless steels, which is the core objective of this work.

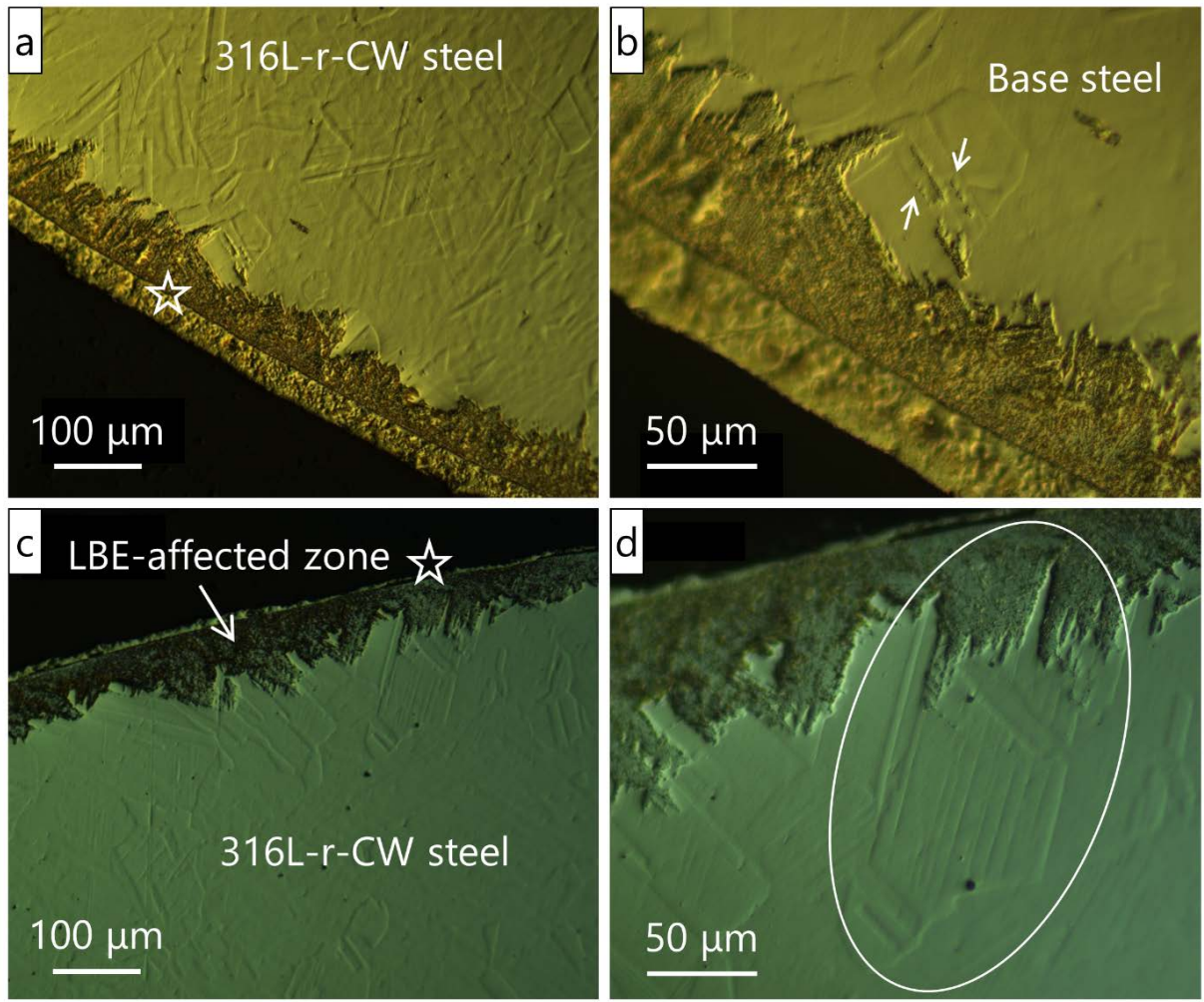

**Fig. 14.** LOM images of OPS-polished 316L-r-CW exposed to oxygen-poor static LBE at 500°C for 1003 h (previous work [8], unpublished data). (a,b) Overview and detail of an area showing LBE penetration along annealing twins (arrows). (c,d) Overview and detail of an area showing a set of closely spaced deformation twins (ellipse) that favours LBE penetration into the steel.

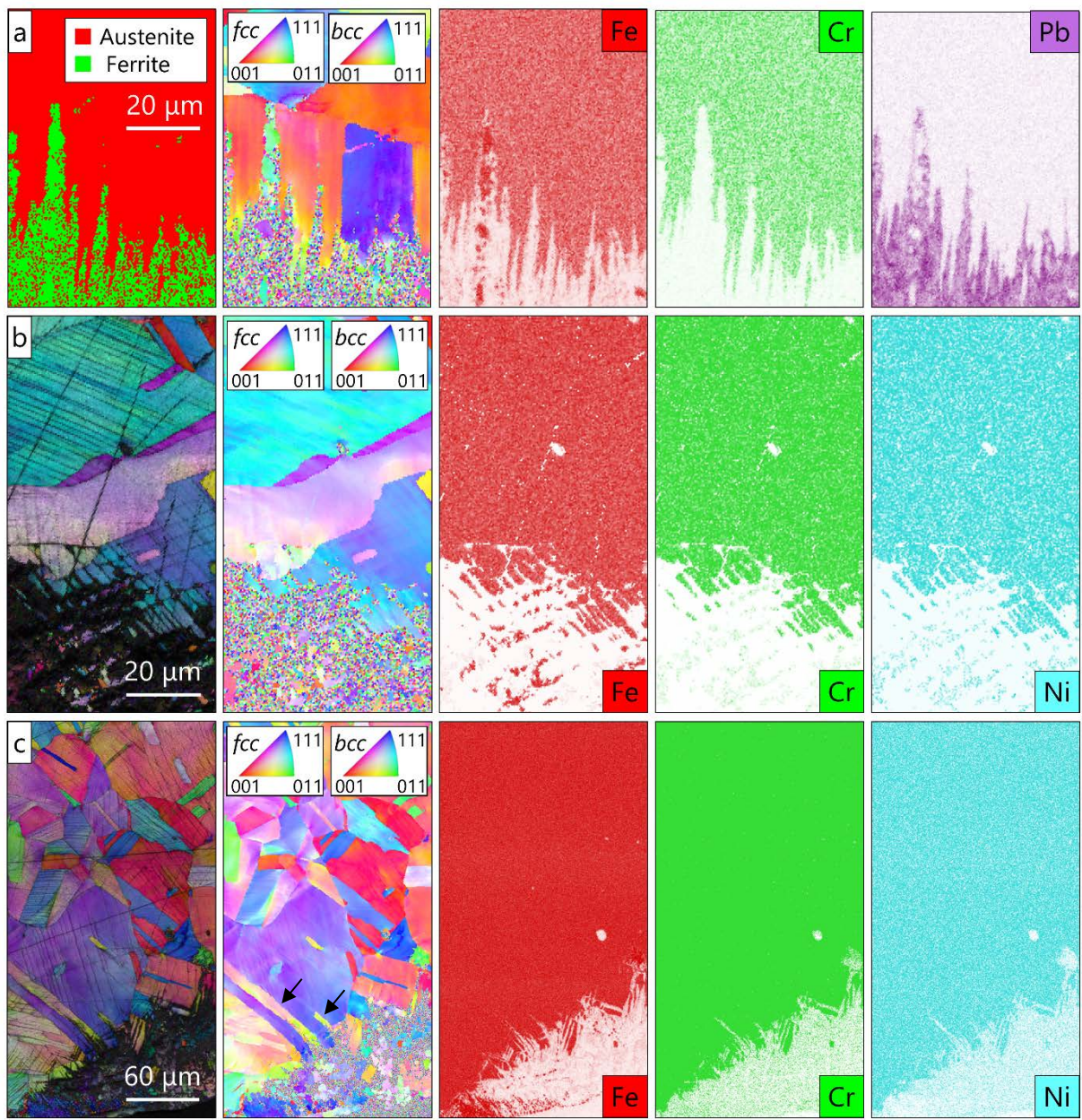

**Fig. 15.** (a) EBSD phase & orientation maps, and EDS elemental maps of the dissolution front in 316L-r-CW; the EDS maps show ferritized (Fe-rich, Cr-depleted) steel grains in the dissolution-affected zone. (b) EBSD orientation map and EDS elemental maps of the dissolution front in 316L-r-CW; the dissolution front advances with the help of ultra-fine deformation twins. (c) EBSD orientation map and EDS elemental maps of the dissolution front in 316L-p-SA; the dissolution front advances with the help of annealing twins (arrows) and much finer deformation twins.

Additional EBSD data showing the interaction of annealing/deformation twin boundaries with the dissolution process in 316L-p-SA and 316L-r-CW exposed to oxygen-poor, static LBE at 500°C for 619h (test 1) are presented in Fig. 15. The 316L-r-CW dissolution front morphology in Fig. 15a is very similar to that shown in Fig. 14d, which is clearly created by the presence of closely-spaced deformation twins. The EBSD phase map of Fig. 15a confirms the ferritisation of the dissolution-affected zone, while the accompanying EDS elemental maps point out that LBE dissolution attack started by selective leaching of the highly soluble steel alloying elements (Ni, Mn and Cr – only the Cr map is provided here due to space limitations), as Fe-containing grains are still visible in the dissolution-affected zone (see Fe map). Fig. 15b shows the interaction of thin deformation twins with the dissolution front in 316L-r-CW, while Fig. 15c shows the combined interaction of thin deformation twins and coarser annealing twins with the dissolution front in 316L-p-SA. Both 316L steel specimens were exposed simultaneously to oxygen-poor static LBE in this work (test 1).

### 3.4.2 Steel precipitates in DIN 1.4970 steels

Steel precipitates (oxides, sulphides, δ-ferrite) were reported by Lambrinou et al. [8] to locally promote LBE dissolution attack in 316L stainless steels, as the steel/precipitate interfaces appear to be preferred paths for the LBE ingress into the steel bulk. Stringer-like precipitates (oxides, δ-ferrite) in cold-rolled stainless steel heats of industrial size, such the 316L-p-SA steel used in this work, are of particular importance, because they allow the LBE ingress over long distances into the steel, as they are usually elongated precipitates of several micrometres in length.

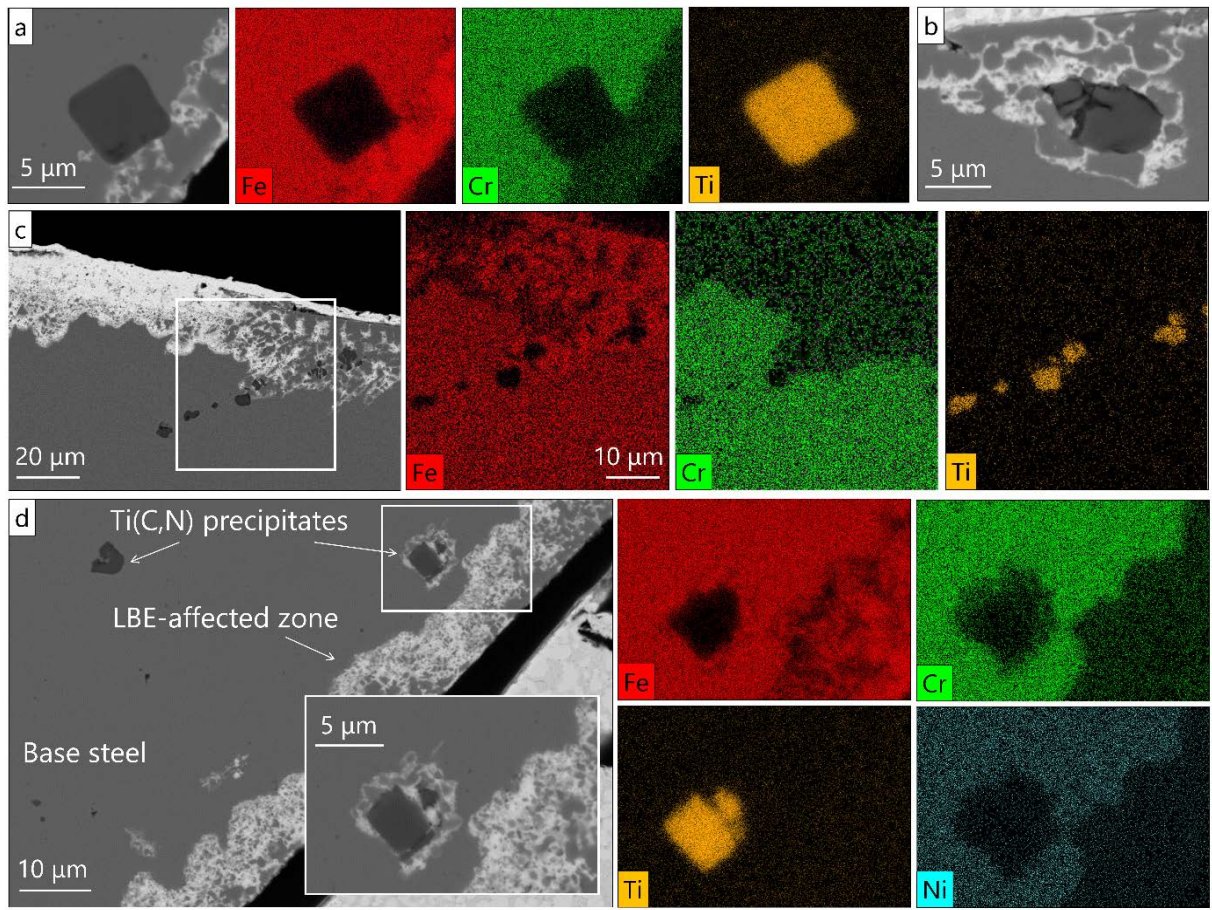

**Fig. 16.** BSE images and EDS elemental maps of Ti(N,C) precipitates in exposed DIN 1.4970. (a) First interaction of a single precipitate with the dissolution front of brushed 1.4970-t-24-CW. (b) Large precipitate advances locally the dissolution front in 1.4970-t-46CW by ~10 μm. (c) Linear arrangement of precipitates extends locally the dissolution-affected zone in 1.4970-r-SA. (d) Precipitates just ahead the main dissolution front favour locally the dissolution process in brushed 1.4970-t-24CW.

This study reports on the effect of the Ti(C,N) precipitates, which are intentionally formed in the DIN 1.4970 steel (see section 3.1), on the steel dissolution corrosion behavior, as there are no previously published data on this issue. Similar to what was previously observed with 316L steel precipitates, the interfaces between the DIN 1.4970 steel matrix and Ti(C,N) precipitates served as paths of preferred LBE ingress into the steel (Fig. 16). As may be seen in the SEM/EDS data of Figs. 16b, 16c and 16d, the Ti(C,N) precipitates promote the local advancement of the dissolution front (the magnitude of that advancement depends on their size). Linear arrangements of Ti(C,N) precipitates (Fig. 16c) that run through the fuel cladding wall thickness are very undesirable steel defects, as they could cause early fuel cladding failure due to LBE dissolution attack. It is important to realize that, since Ti(C,N) precipitates cannot be oxidized, the presence of large precipitates at the steel surface could easily lead to dissolution 'pitting', as LBE will surely penetrate through the steel/precipitate interface into the steel, even when the rest of the steel surface is covered by a protective oxide scale. Hence, the careful steel manufacture is of essence for fuel cladding tubes intended for use in Gen-IV LFRs, as the presence of large Ti(C,N) precipitates close to the steel surface must be avoided to the extent possible.

3.5     Aspects of DIN 1.4970 steel ferritisation

3.5.1   Selective leaching and ferritisation

The ferritisation of 316L austenitic stainless steels caused by selective leaching (i.e., removal of the austenite stabilisers Ni and Mn, due to LBE dissolution attack) has already been understood and discussed in previous studies [8,19]. This work expands the current understanding of this undesirable LMC effect on the MYRRHA reference DIN 1.4970 fuel cladding steel, thus addressing ferritisation for both MYRRHA candidate steels (i.e., the 316L structural steel and the DIN 1.4970 fuel cladding steel). The ferritisation of both reference 1.4970-t-24CW and annealed 1.4970-t-HT was studied by means of SEM/EDS, EBSD and FIB/TEM. Figs. 17a-17b and Figs. 18a-18b provide evidence of early-stage dissolution-induced ferritisation in 1.4970-t-24CW and 1.4970-t-HT, respectively, by means of SEM/EDS and EBSD; in this steel area, the dissolution-affected zone is confined within a sub-surface layer of grains (2.5-4.0 μm). FIB was used to lift-out thin foils from the dissolution-affected zone for TEM analysis; due to its sensitivity to ion beam milling, LBE was preferentially removed from some parts of the thin foil, leaving holes behind (Fig. 17c). Selected area diffraction patterns of both the unaffected steel bulk and the dissolution-affected zone (Figs. 17d and 17e, respectively) confirmed beyond any doubt the *fcc*-to-*bcc* phase transformation, due to selective leaching of the steel in contact with oxygen-poor liquid LBE. EDS analysis of the same foil (not shown here) verified that the dissolution-affected zone suffered selective leaching of Ni, Mn and Cr. Moreover, no gradients in the concentration of these steel alloying elements was found close to the austenite/ferrite interface. The BF TEM image of Fig. 17c shows that the austenite (*fcc*) phase has a high density of forest dislocations due to the steel (24%) cold work, while the derivative ferrite (*bcc*) phase appears to be defect-free and with a smaller grain size than the 'parent' austenite phase.

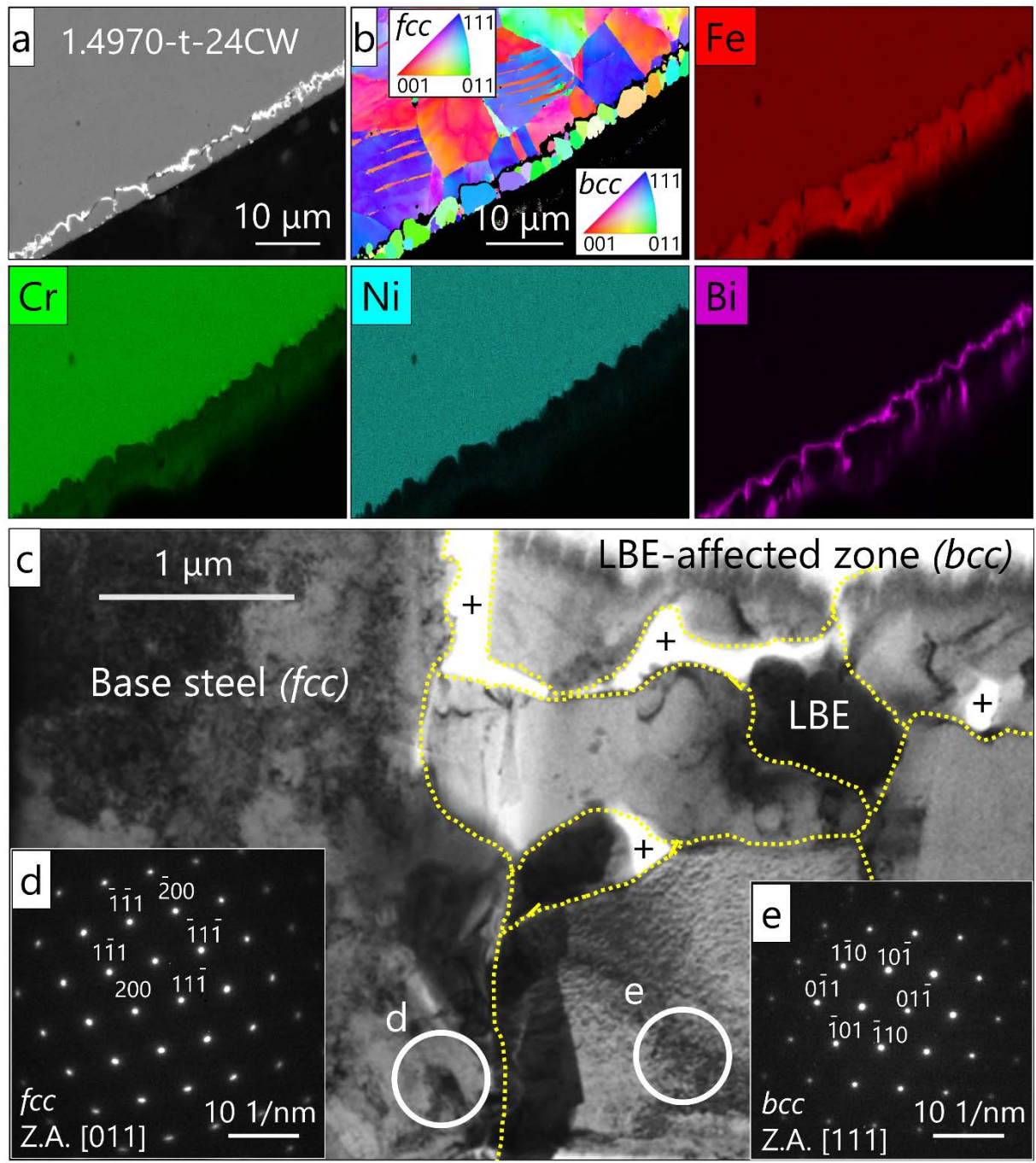

**Fig. 17.** Early-stage dissolution-induced ferritisation in 1.4970-t-24CW (test 3). (a) BSE image, and (b) EBSD orientation map and EDS elemental maps showing the ferritisation of the dissolution-affected zone, which is confined in a layer of grains just below the steel surface. (c) TEM BF image of the ferrite/austenite interface, and (d,e) SADPs of the circled areas in Fig. 17c confirming the ferritisation (area e) of the austenitic (area d) steel. The + symbol shows voids in the thin foil due to LBE removal.

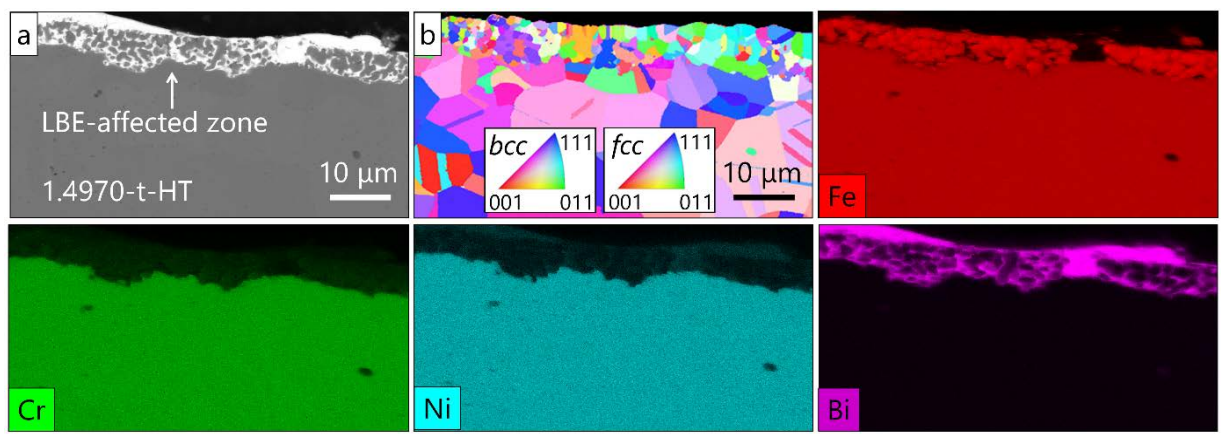

**Fig. 18.** Ferritisation of the dissolution-affected zone in 1.4970-t-HT (test 3). (a) BSE image, and (b) EBSD orientation map and EDS elemental maps of the dissolution-affected area.

*3.5.2   Austenite/ferrite crystallographic orientation relationship*

Charalampopoulou et al. [52] reported the existence of a specific orientation relationship (OR) between the original austenite phase (γ; *fcc*) and the derivative ferrite (α; *bcc*) phase involved in the γ → α phase transformation, also known as 'ferritisation', observed in two 316L steel heats exposed to oxygen-poor ($C_0 \approx 10^{-9}$ mass), static liquid LBE at 550°C for 1000 h. The 316L steel heats used in that earlier study were one solution-annealed (316L-SA; this heat is here named 316L-p-SA), and one cold-worked (316L-CW; ⌀ 10 mm cylindrical rod produced by Panchmahal Steel Ltd., India). As already mentioned, ferritisation of austenitic stainless steels results from the selective leaching of the highly soluble steel alloying elements Ni and Mn, which are the austenite phase stabilisers. Typically, *fcc*-to-*bcc* phase transformations in austenitic stainless steels obey the Kurdjumov-Sachs or Nishiyama-Wassermann OR models. However, the earlier study on 316L steels showed that the *fcc*-to-*bcc* phase transformation due to LBE dissolution corrosion-induced ferritisation obeyed the Pitsch OR model [52].

In this work, the γ → α phase transformation was studied in DIN 1.4970 austenitic stainless steels in order to check whether differences in the steel chemical composition and microstructure affect the austenite/ferrite OR. DIN 1.4970 fuel cladding tubes 1.4970-t-24CW (as-received) and 1.4970-t-HT (annealed at 1000°C for 2 h) were exposed to oxygen-poor, static LBE at high temperatures (T ≥ 500°C) for 600-1000 h, thereby undergoing ferritisation of the dissolution-affected zone. EBSD orientation maps were taken from numerous locations on the metallographic cross-sections of the two DIN 1.4970 cladding tubes, so as to ensure that the population of acquired data was statistically representative of the γ → α transformation. Grain misorientations at the γ/α interface were characterised using both misorientation angles and orientation density functions (ODFs) derived by EBSD, as shown in Figs. 19 and 20, respectively. Once again, due to the presence of liquid LBE at the ferritisation front, the *fcc*-to-*bcc* phase transformation was best described by the Pitsch OR model, which has been previously described as a fundamental lattice distortion in steels undergoing similar phase transformations [52]. The deeper understanding of this type of γ → α phase transformation, for the first time studied by Charalampopoulou et al. [52] in 316L stainless steels affected by LBE dissolution attack, is a first step towards its possible mitigation. Possible mitigation approaches, such as closely controlled cold working steps, might be able to induce sub-surface crystallographic textures that are more resistant to LBE dissolution (e.g., we saw in section 3.1 that strong <111> textures parallel to the loading direction during cold deformation enhance the steel susceptibility to dissolution corrosion due to the high density of deformation twins in these grains). Such steels, provided that their fabrication is possible on an industrial scale, would be better suited for use in Gen-IV LFRs.

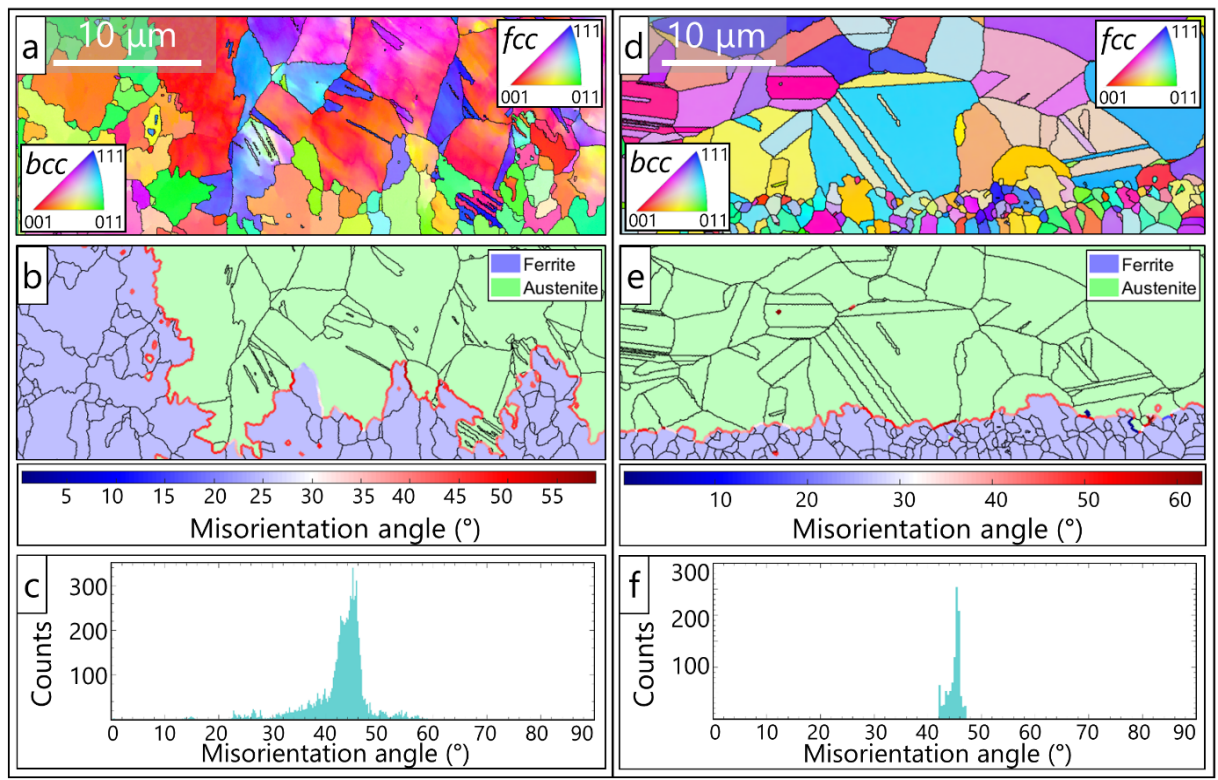

**Fig. 19.** EBSD orientation maps of the ferritized dissolution-affected zone in (a) 1.4970-t-24CW and (d) 1.4970-t-HT, with inverse pole figure colour maps for the *fcc* and *bcc* phases. Corresponding EBSD phase maps showing the γ/α interface misorientation boundary in (b) 1.4970-t-24CW and (e) 1.4970-t-HT. Misorientation angle distributions at the γ/α interface misorientation boundary in (c) 1.4970-t-24CW and (f) 1.4970-t-HT.

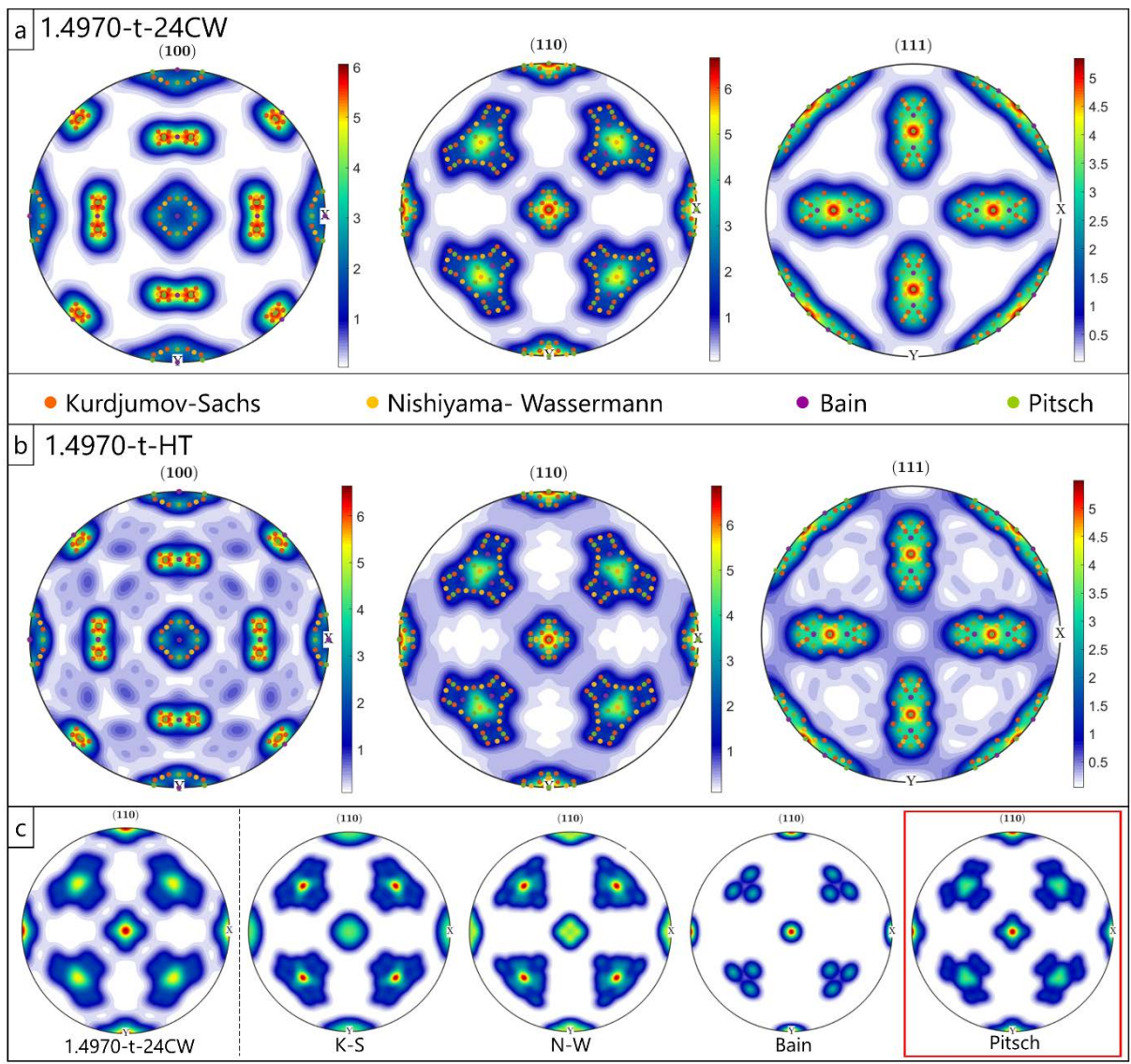

**Fig. 20.** Iso-density pole figures of the experimental data acquired from (a) 1.4970-t-24CW and (b) 1.4970-t-HT, plotted with the theoretical pole arrangements of the four OR models (colour bar in multiples of random distribution unit). (c) Comparison of estimated iso-density pole figures from experimental data with simulated ones, for the different theoretical ORs in (110) pole figures.

### 3.6 Anticorrosion coatings on DIN 1.4970 fuel cladding tubes

#### 3.6.1 Microstructural characterization of pristine anticorrosion coatings

The $Cr_2AlC$ MAX phase coating had a rather uniform thickness of ~3 μm and is characterized by a columnar microstructure (Fig. 21a), typical for $Cr_2AlC$ coatings deposited by magnetron sputtering [53,54]. The successful formation of the $Cr_2AlC$ MAX phase has been confirmed by X-ray diffraction (XRD; not shown here), and the coating adhered well on the 1.4970-t-24CW substrate fuel clad. Coating inspection by SEM/EDS revealed a Ni-enriched and Cr-impoverished thin (<300 nm) layer at the interface between the $Cr_2AlC$ coating and the DIN 1.4970 substrate fuel clad. It is reasonable to assume that the relatively high deposition temperature (580°C) might have caused interdiffusion of elements between steel and coating at the steel/$Cr_2AlC$ interface.

This is the first time, to the authors' knowledge, that a Cr$_2$AlC MAX phase-coated DIN 1.4970 fuel cladding steel has been exposed to liquid LBE, so as to assess its protectiveness against dissolution corrosion.

The attempts to produce a V$_2$AlC coating on the 1.4970-t-24CW fuel clad by means of cathodic arc deposition proved unsuccessful, as the deposition temperature (500-550°C) was too low to ensure the formation of the V$_2$AlC MAX phase. The deposited thin (<200 nm) film (Fig. 21c) was a defective V-based carbide layer with the V$_2$Al$_x$C$_y$ general stoichiometry due to the uncertainty in its Al and C contents. This thin film was characterised by a non-optimal adhesion on the 1.4970-t-24CW substrate fuel clad, covering roughly one fourth of the surface of the substrate. XRD analysis (not shown here) of the V$_2$Al$_x$C$_y$ thin film showed that it consisted mostly of Al-containing V$_2$C (orthorhombic, ortho). Despite its imperfect state, it was decided to expose the V$_2$Al$_x$C$_y$-coated DIN 1.4970 fuel clad to liquid LBE, so as to check whether the presence of a thin carbide layer on the steel surface would somehow affect the resistance of the substrate clad to dissolution corrosion.

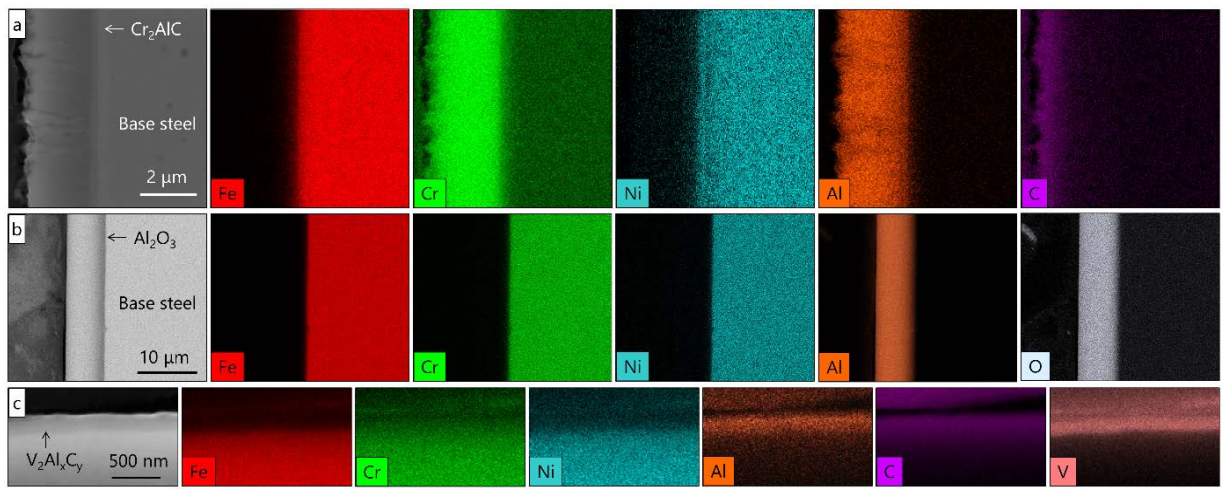

**Fig. 21.** SEM images and EDS elemental maps of as-deposited (a) Cr$_2$AlC MAX phase, (b) Al$_2$O$_3$, and (c) V$_2$Al$_x$C$_y$ coatings on 1.4970-t-24CW.

The Al$_2$O$_3$ coating deposited by PLD was compact, defect-free, uniform in thickness (~7 μm), and well-adherent to the 1.4970-t-24CW substrate fuel clad (Fig. 21b). In accordance with prior experience on PLD-deposited Al$_2$O$_3$ coatings [15-19], the as-deposited coating was amorphous, since the only detectable sharp peaks in the acquired XRD patterns (not included) corresponded to the DIN 1.4970 steel substrate. TEM analysis of the Al$_2$O$_3$ coatings confirmed the absence of nanoscale porosity or other defects, showing a low density of nanocrystalline domains (average size: 2-10 nm) that were homogeneously dispersed in an amorphous matrix, as reported earlier [18,55]. This particular microstructure is responsible for the appealing properties of these coatings, including a superior hardness that accounts for high resistance to erosion, scratching or fretting [17,18]. Recently, in-situ nanomechanical testing of free-standing Al$_2$O$_3$ thin (50 nm-thick) layers demonstrated a remarkable elastoplastic response under both tensile and compressive loading at RT, attesting a deformability up to ~15% [56]; such deformability is desirable for fuel cladding coating materials, as fuel clads tend to deform in-service, primarily due to irradiation creep. In line with their already established technological maturity, PLD-grown $Al_2O_3$ coatings on steel substrates have been previously exposed up to 8000 h to both static liquid Pb and Pb-16Li, exhibiting superior protectiveness against undesirable LMC effects [16,18,57,58].

### 3.6.2 Mitigation of dissolution corrosion by anticorrosion coatings

Metallographic cross-sections of both uncoated 1.4970-t-24CW (brushed & polished surface state) and coated 1.4970-t-24CW with the $Cr_2AlC$ MAX phase, $Al_2O_3$, and $V_2Al_xC_y$ were studied by SEM/EDS to assess the coatings' protectiveness against dissolution corrosion of the substrate clad; these specimens have been previously exposed to oxygen-poor, static LBE at 500°C for 1000 h (test 4). As expected, the uncoated 1.4970-t-24CW was affected by dissolution corrosion on both its inner and outer surfaces (Fig. 22a). Typically, LBE dissolution attack on both tube surfaces started intergranularly (Fig. 22b; see BSE images of the inner surfaces of $Cr_2AlC$- and $Al_2O_3$-coated 1.4970-t-24CW) and proceeded transgranularly (Fig. 22a), esp. in areas where the dissolution corrosion damages were deep; as shown in Fig. 8b, the maximum dissolution depth in uncoated 1.4970-t-24CW was ~27 μm. Very often, a layer of sub-surface ferritized grains (<5 μm in thickness) remained on top of the dissolution-affected zone (Fig. 22a), which appeared to have been attacked transgranularly; the severity of attack was not uniform throughout the dissolution-affected zone, and in some areas even the ferritized (Fe-rich) grains were almost gone.

Both $Cr_2AlC$ and $Al_2O_3$ coatings remained protective to the 1.4970-t-24CW substrate under the exposure conditions of test 4 (500°C, $C_O$ < 5×10$^{-11}$ mass%, 1000 h). The columnar microstructure of the $Cr_2AlC$ MAX phase coating caused some initial anxiety with respect to the possible LBE penetration between the columnar $Cr_2AlC$ grains during the performed exposure. However, no LBE was detected in the intercolumnar spaces (Fig. 22d); even though this is a very promising first result for the possible use of MAX phase coatings for dissolution corrosion mitigation in Gen-IV LFR fuel cladding steels, longer-term exposures are required to assess the true potential of this coating technology. It is worthwhile mentioning that the thickness of the Ni-enriched layer at the steel/$Cr_2AlC$ interface has grown to ~500 nm, while another Ni-impoverished layer of ~1 μm in thickness has developed into the steel (Fig. 22d). This suggests that the Ni-enriched thin layer at the interface of as-deposited $Cr_2AlC$ with the steel substrate belonged to the bottom coating part; this coating part is expected to be composed by ultra-fine, equiaxed (due to the maximum cooling rate upon coating deposition) and Ni-containing $Cr_2AlC$ grains. The basis of this assumption lies in the knowledge that outward diffusion of steel alloying elements, such as Ni, is expected to occur during the prolonged (1000 h) steel exposure to LBE at 500°C (test 4). It is, therefore, likely that Ni (smallest atomic radius: 124 pm; atomic radii of Cr and Fe: 128 pm and 126 pm, respectively) diffuses into the $Cr_2AlC$ grains, forming in-situ a MAX phase solid solution. Future TEM analysis of both as-deposited and exposed Cr$_2$AlC coating is needed to show whether these assumptions are correct; however, the validity of assumptions fall outside the scope of this work, which assesses the relative protectiveness of candidate anticorrosion coatings against dissolution corrosion. The protectiveness of the Al$_2$O$_3$ coating was excellent, as expected from previous exposures of PLD-deposited Al$_2$O$_3$ coating in liquid Pb and Pb-16Li [16,18,57,58]. The low deposition temperature (RT) of the Al$_2$O$_3$ coating did not cause any compositional modifications close to the steel surface. Since all DIN 1.4970 fuel cladding tubes were immersed open to the LBE bath, the inner steel surface was invariably affected by dissolution corrosion, as shown for the Cr$_2$AlC- and Al$_2$O$_3$-coated 1.4970-t-24CW in Figs. 22b-22c.

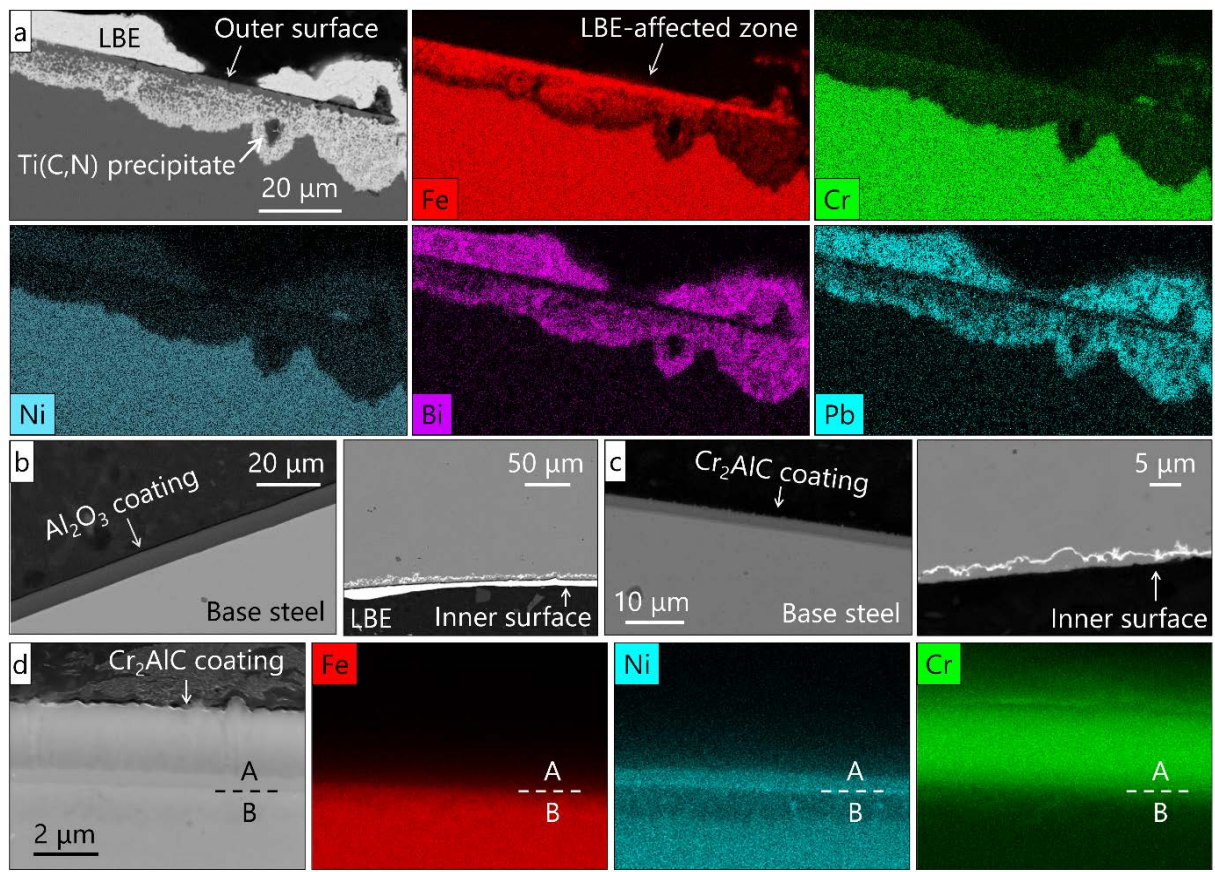

**Fig. 22.** (a) BSE image and EDS elemental maps of the dissolution-affected zone in uncoated 1.4970-t-24CW exposed to LBE for 1000 h in this work. BSE images of the outer tube surfaces in 1.4970-t-24CW coated with (b) Al$_2$O$_3$ and (c,d) Cr$_2$AlC after LBE exposure. BSE images of the inner (uncoated) surfaces of the Al$_2$O$_3$- and Cr$_2$AlC-coated tubes are also shown for comparison in (b) and (c), respectively. The bottom part of the Cr$_2$AlC coating appears to be enriched in Ni, while the top steel layer (~1 μm) is Ni-impoverished; the interface between Cr$_2$AlC coating (A) and steel (B) is indicated by a dashed line.

The thinness (<200 nm) and partial (1/4 of the cladding tube surface) coverage of the V$_2$Al$_x$C$_y$ coating did not allow the reliable evaluation of its protectiveness, as it was unclear after the LBE exposure which DIN 1.4970 fuel cladding tube areas were covered by the coating and which ones not. Detailed and labor-intensive TEM work might be required to analyze areas where thin film residues are still present, however, this was deemed ill-advised considering that this coating did not have the adequate quality to justify an extensive post-exposure analysis. Still, no visible LBE dissolution attack of the outer steel surface was detected by SEM, which suggests that even thin V-Al-C films might be capable of improving the steel's resistance to dissolution corrosion. The imperfect state of the herein exposed $V_2Al_xC_y$ coating does not permit further speculations. One should also consider that cathodic arc might not be the best method to deposit $V_2AlC$; for example, Azina et al. [59] have recently used magnetron sputtering to deposit $V_2AlC$ MAX phase coatings of uniform thickness (~4 μm) and columnar microstructure on MgO(100) substrates. The $V_2AlC$ deposition temperature was 580°C, which allows the transfer of this technology on DIN 1.4970 fuel cladding tubes at a limited risk of property degradation due to the mild deposition conditions.

## 4 Conclusions

In this work, 316L and DIN 1.4970 austenitic stainless steels were exposed to oxygen-poor ($C_0$ < $10^{-8}$ mass%), static liquid LBE at 500°C for 600-1000 h. These two austenitic stainless steels are the MYRRHA candidate steels: the 316L steel is intended for structural components, while the DIN 1.4970 is the selected fuel cladding steel. The main objective of this study was to investigate the early stages in the dissolution corrosion behavior of these two steels, and to assess their relative resistance to dissolution corrosion by simultaneously exposing them to liquid LBE in the uncoated state. Another objective of this study was to compare the protectiveness of select ($Cr_2AlC$, $Al_2O_3$, $V_2Al_xC_y$) anticorrosion coatings on DIN 1.4970 fuel cladding tubes during a 1000 h-long test. The intentional exposure of steels with different grain size, density of defects such as annealing and deformation twins, degree of cold work, surface state, etc., provided valuable insights into the factors affecting the steel dissolution corrosion behavior. The main findings of this study may be summarized as follows:

1. LBE dissolution attack started invariably intergranularly, as GBs are preferred paths for the LBE ingress into the steel bulk. The LBE penetration into GBs was typically followed by the selective leaching of the highly soluble in LBE steel alloying elements (Ni, Mn and Cr). The selective leaching of the austenite stabilizers Ni and Mn resulted in ferritisation (i.e., *fcc*-to-*bcc* phase transformation) of the dissolution-affected zone. It was found that the orientation relationship between the starting austenite (*fcc*) and the derivative ferrite (*bcc*) phases obeyed the Pitsch OR model, in both DIN 1.4970 and 316L stainless steels.

2. The dissolution mode often changed from intergranular to transgranular, as the depth of attack increased. This was associated with the thick diffusion boundary layers established at the steel/LBE interfaces during steel exposures to truly static LBE conditions in this work. Transgranular dissolution typically refers to the complete consumption of the (Fe-based) ferritized grains in the dissolution-affected zone; for industrial-size steel heats, such as the solution-annealed 316L-p-SA, the alternation of zones of intergranular attack with zones of transgranular attack can be associated with the alternation of "bands" of different chemical composition (esp. Ni), a phenomenon that is often observed upon steel cold rolling.

3. A strong interplay between the steel microstructure and the dissolution corrosion process was observed in both steels, whereby LBE penetrated into the steel via preferred paths, such as grain boundaries, annealing/deformation twin boundaries, and interfaces between steel matrix and secondary precipitates, such as the Ti(C,N) precipitates in DIN 1.4970.

4. The most severely affected steels were either coarse-grained steels with high fractions of annealing twins (e.g., 1.4970-r-SA, 316L-r-CW) or fine-grained steels with high fractions of deformations twins (e.g., 316L-t-46CW). Coarse-grained steels with high fractions of annealing twins (e.g., 316L-r-CW) proved to be very susceptible to dissolution 'pitting', esp. when large grains with annealing twins were unfavorably oriented relative to the steel surface. Dissolution 'pitting' is highly undesirable for thin-walled components, such as fuel cladding tubes and heat exchanger tubes. Hence, the steel optimization for a specific reactor component demands the careful consideration of microstructural aspects, such as grain size and density of dissolution-promoting defects (e.g., annealing/deformation twins).

5. Mild differences in surface roughness (e.g., 'brushed & polished' vs. 'brushed' surface states in DIN 1.4970 fuel clads) did not affect the extent of dissolution corrosion damages. Very rough steel surfaces, however, are expected to affect the dissolution corrosion damages, due to the unpredictability of the LMC effects inside deep surface undulations.

6. The effect of texture on the overall steel dissolution corrosion behavior did not appear to be strong. Textures, esp. those with a strong <111> component parallel to the direction of applied load during steel cold deformation, are expected to increase the steel susceptibility to dissolution corrosion, due to the increased density of deformation twins in grains with that crystallographic orientation parallel to the direction of applied load. However, the effect of grains with a particular crystal orientations on the advancement of the dissolution front is primarily local. However, one cannot exclude the possibility that a strict control of the steel texture might exploit the existing austenite/ferrite orientation relationship in both 316L and DIN 1.4970 steels, thereby limiting the extent of dissolution-induced ferritisation.

7. High-temperature annealing (1000°C, 2 h) of DIN 1.4970 fuel cladding tubes did not affect the severity of dissolution damages, despite the steel grain coarsening and steep increase in the fraction of annealing twins. This is attributed to the fact that the coarsened grains were still not large enough to result in severe dissolution corrosion effects, such as 'pitting'.

8. The deposition of select anticorrosion coatings (i.e., $Cr_2AlC$, $Al_2O_3$, $V_2Al_xC_y$) on DIN 1.4970 fuel cladding tubes proved effective means of dissolution corrosion mitigation, at least for the specific test conditions (500°C, $C_0$ < 5×10$^{-11}$ mass%, 1000 h) used in this work.

**Acknowledgements**

The authors gratefully acknowledge the support (in terms of test setups and supply of 316L & DIN 1.4970 steel specimens) provided in the framework of the MYRRHA project. The fabrication of anticorrosion coatings was funded by the Euratom research and training programme 2014-2018 under Grant Agreement No. 740415 (H2020 IL TROVATORE) and Grant Agreement No. 755269 (H2020 GEMMA). E. Charalampopoulou and K. Lambrinou would like to personally thank J. Joris for technical support during the launching and follow up of all corrosion tests, D. Penneman for his help with the annealing heat treatment of the 1.4970-t-24CW tube segment, as well as J. Lim for the manufacture and calibration of the oxygen sensors used in this study. The research leading to these results falls within the framework of the European Energy Research Alliance Joint Programme on Nuclear Materials (EERA JPNM).